\documentclass[10pt, conference, letterpaper]{IEEEtran}\IEEEoverridecommandlockouts
\usepackage{cite}
\usepackage{amsmath,amssymb,amsfonts}
\usepackage{algorithmic}
\usepackage{graphicx}
\usepackage{textcomp}
\usepackage{xcolor}
\def\BibTeX{{\rm B\kern-.05em{\sc i\kern-.025em b}\kern-.08em
    T\kern-.1667em\lower.7ex\hbox{E}\kern-.125emX}}
    
\usepackage{verbatim}

\usepackage{amsmath,amssymb,amsthm,nicefrac}
\usepackage{color,array}
\usepackage{xspace}
\usepackage{dsfont}
\usepackage{hyperref}
\usepackage{listings}             
\usepackage{subcaption}
\usepackage{multirow}

\newcommand{\id}{\ensuremath{\mathds{1}}}
\newcommand{\reals}{\ensuremath{\mathbb{R}}}
\newcommand{\naturals}{\ensuremath{\mathbb{N}}}
\newcommand{\prob}{\ensuremath{\mathbf{P}}}
\newcommand{\expect}{\ensuremath{\mathbb{E}}}
\newcommand{\catalog}{\mathcal{C}}

\newcommand{\capacity}{\ensuremath{c}}
\newcommand{\source}{\ensuremath{\mathcal{S}}}
\newcommand{\pathset}{\ensuremath{\mathcal{P}}}
\newcommand{\ppath}{\ensuremath{p}}
\newcommand{\requests}{\ensuremath{\mathcal{R}}}
\newcommand{\feasibledomain}{\mathcal{D}}
\newcommand{\argmax}{\mathop{\arg\max\,}}

\usepackage[ruled]{algorithm2e}

\newcommand{\arriveprob}{\lambda}
\newcommand{\cachestate}{x}
\newcommand{\cachestatevec}{\boldsymbol{\cachestate}}

\newcommand{\cacheprob}{\xi}
\newcommand{\cacheprobvec}{\boldsymbol{\cacheprob}}

\newcommand{\route}{r}
\newcommand{\routevec}{\boldsymbol{\route}}

\newcommand{\routeprob}{\rho}
\newcommand{\routeprobvec}{\boldsymbol{\routeprob}}

\newcommand{\dual}{\psi}
\newcommand{\dualvec}{\boldsymbol{\dual}}

\newcommand{\jvar}{\boldsymbol{y}}

\newtheorem{theorem}{Theorem}
\newtheorem{lemma}{Lemma}

\newcounter{packednmbr}

\newcommand{\SR}{\ensuremath{\mathtt{TOT}}}

\usepackage[font=small]{caption}
\newcommand{\notes}[1]{\textcolor{black}{#1}}
\newcommand{\final}[1]{\textcolor{black}{#1}}

\newcommand{\arxiv}[2]{#1} 
\usepackage{setspace}

\begin{document}

\title{Jointly Optimal Routing and Caching \\
with Bounded Link Capacities
\thanks{\final{The authors gratefully acknowledge support from National Science Foundation grants NeTS-1718355 and  CCF-1750539, National Natural Science Foundation of China under Grant 62172054 and the National Key R\&D Program of China under Grant 2019YFB1802603.}}
}

\author{\IEEEauthorblockN{ Yuanyuan Li,\textsuperscript{1} Yuchao Zhang,\textsuperscript{3,2} Stratis Ioannidis,\textsuperscript{1} and Jon Crowcroft\textsuperscript{2}}
\IEEEauthorblockA{ \textsuperscript{1}Northeastern University, \textsuperscript{2}University of Cambridge, \textsuperscript{3}Beijing University of Posts and Telecommunications  \\
Email: yuanyuanli@ece.neu.edu, yczhang@bupt.edu.cn, ioannidis@ece.neu.edu, jon.crowcroft@cl.cam.ac.uk
}
}

\maketitle

\begin{abstract}
We study a cache network in which intermediate nodes equipped with caches can serve requests. We model the problem of jointly optimizing caching and routing decisions with link capacity constraints over an arbitrary network topology. This problem can be formulated as a continuous \final{diminishing-returns }(DR)-submodular maximization problem under multiple continuous DR-supermodular constraints, and is NP-hard. We propose a poly-time alternating primal-dual  heuristic algorithm, in which primal steps produce solutions within $1-\frac{1}{e}$ approximation factor from the optimal. Through extensive experiments, we demonstrate that our proposed algorithm significantly outperforms competitors.
\end{abstract}

\arxiv{
\begin{IEEEkeywords}
cache networks, DR-submodular, Lagrangian, primal dual, Frank Wolfe
\end{IEEEkeywords}
}{}

\section{Introduction}
\label{sec:intro}
The problem of optimally storing content  in a network arises in  a broad array of networking applications and systems, including information-centric networks (ICNs)\arxiv{~\cite{jacobson2009networking,yeh2014vip}}{~\cite{wang2016fair}}, content-delivery networks (CDNs)\arxiv{~\cite{orchestrating2012,dehghan2014complexity}}{~\cite{dehghan2014optimal}}, \arxiv{}{and}  wireless/femtocell networks\arxiv{~\cite{shanmugam2013femtocaching,naveen2015interaction,poularakis2013approximation}}{~\cite{shanmugam2013femtocaching}}, \arxiv{web-cache design~\cite{laoutaris2004meta,che2002hierarchical,zhou2004second}, and peer-to-peer network~\cite{cohen2002replication,ioannidis2009absence},} to name a few. It has recently been the focus of several studies that aim to design \emph{cache networks} with optimality guarantees \cite{wang2013optimal,dehghan2014optimal,ioannidis2016adaptive,ioannidis2018jointly,shanmugam2013femtocaching,wang2016fair,li2020universally}. 
Such works optimize either caching decisions alone \cite{wang2013optimal,dehghan2014optimal, ioannidis2016adaptive} or caching and routing jointly \cite{dehghan2014optimal,ioannidis2018jointly}. Objectives include, e.g., minimizing aggregate transfer costs \cite{ioannidis2016adaptive, wang2013optimal} or queuing delays \cite{mahdian2020kelly,li2020universally,li2021cache}, maximizing a fairness objective \cite{wang2016fair,liu2021fair} or throughput\cite{liu2019joint,kamran2021rate}, etc. 
 
Following Ioannidis and Yeh~\cite{ioannidis2018jointly}, \arxiv{we consider a network in which a fixed set of servers store content permanently. Nodes corresponding to customer-facing gateways generate requests and choose among several different routes to send requests to these servers. Intermediate, cache-enabled nodes, corresponding to storage-augmented  routers over the path, can also store contents, and immediately serve requests for content they store.}{we consider a network in which requests for content generated by customer-facing gateways are routed towards fixed servers, but can be served by intermediate, cache-enabled nodes.} The network designer's goal is to determine (a) how to route requests, as well as (b) where to place contents, to minimize overall transfer costs. Even though this problem is NP-hard, Ioannidis and Yeh  \cite{ioannidis2018jointly} provide a polytime approximation algorithm, and show that joint optimization of caching and routing decisions can reduce transfer costs by three orders of magnitude, in practice.

This analysis  assumes infinite link capacities, which is implausible for real-life networks. We depart by introducing \emph{link capacity constraints}: we assume every edge in the network can carry at most a constant amount of traffic per second. This is clearly more realistic, but also leads to optimization problems of a vastly different nature\arxiv{ than the ones considered by Ioannidis and Yeh}{ than \cite{ioannidis2018jointly}}.
For example, unbounded capacities result in deterministic optimal solutions, whereby each demand is routed over a single, unique path. In contrast, introducing link capacity constraints gives rise to \emph{multi-path} optimal solutions: optimal traffic may split across multiple routes. From a technical standpoint, introducing link capacities drastically changes our optimization problem. In contrast to the vast majority of prior research in the area \cite{shanmugam2013femtocaching, mahdian2020kelly, ioannidis2018jointly}, our constraints no longer form a matroid; this requires a very different algorithm than the one employed by~\cite{ioannidis2018jointly}.

\arxiv{}{\final{Several works \cite{dehghan2015complexity,poularakis2020service, poularakis2013approximation} consider congestion control in bipartite (one-hop) cache networks, but their network model and guarantees do not apply to arbitrary (multi-hop) topologies. Closer to us, Liu et al.~\cite{liu2019joint} and Kamran et al. \cite{kamran2021rate} provide approximation guarantees under link capacity constraints in arbitrary topologies, but differ in both their objective (throughput maximization) and constraints. In particular, there is no notion of routing costs, as incorporated in our setting. Moreover, none of above works gives rise to the DR-submodular structures we observe in our optimization problem. Overall,  these existing algorithms cannot be applied to our setting.}} Our contributions are as follows:
\begin{itemize}
	 \item We model the problem of joint optimization of caching and routing decisions with link capacity constraints over an arbitrary topology. Our model yields  a continuous DR-submodular maximization problem under a set of continuous DR-supermodular constraints.
	 \item The objective is not concave and constraints are not convex. We propose a polynomial-time Lagrangian primal-dual algorithm for this problem. Though the combined, end-to-end algorithm is a heuristic, we show that a $1-\frac{1}{e}$ approximation guarantee holds during primal steps.
	 \item Finally, we conduct extensive experiments over both synthetic and trace-driven networks: our proposed algorithm outperforms several baselines significantly w.r.t. both cache gain and feasibility.
 \end{itemize}

The remainder of this paper is organized as follows. \arxiv{In Sec.~\ref{sec:related}, we review related work.}{} Sec.~\ref{sec:model} introduces the model of cache networks and formulates joint \arxiv{caching and routing }{}optimization problem\arxiv{ with both cache and link capacity constraints}{}. Sec.~\ref{sec:main results} describes our analysis of the problem and proposed algorithm. We present numerical experiments in Sec.~\ref{sec:experiments} and conclude in Sec.~\ref{sec:conclusion}. \arxiv{}{The extended version of this paper is available in \cite{li2022extended}.}

\arxiv{
\section{Related Work}
\label{sec:related}
\noindent\textbf{Optimization Objectives.} 
Several works assign a constant cost to each edge in the network, and aim at making caching decisions that minimize expected routing costs. This objective has been studied in the context of femtocaching systems \cite{shanmugam2013femtocaching}, arbitrary cache networks \cite{ioannidis2016adaptive, ioannidis2018jointly},  small cell networks \cite{hajri2017energy},  parallel computing frameworks \cite{yang2018intermediate} and in proactive (i.e., predictive) cache networks \cite{shukla2017proactive}, to name a few. 
Content placements that maximize the number of requests served by caches are studied in hierarchical caching networks \cite{poularakis2016complexity}, in cellular networks with moving users \cite{poularakis2016code}, in arbitrary congestible networks \cite{liu2019joint, kamran2021rate}, and in multi-cell mobile edge computing networks with storage, computation, and communication constraints \cite{poularakis2020service}. 

To minimize the expected delay experienced by all the requester, Domingues et al. \cite{domingues2017enabling} study the interplay between content search and content placement and Poularakis et al. \cite{poularakis2018distributed} study the content placement of layered-video. Yeh et al. \cite{yeh2014vip} focus on maximizing throughput, i.e., user demand rate satisfied by the network. 
Li. et al. \cite{li2020universally,li2021cache} and Mahdian et al. \cite{mahdian2020kelly} minimize non-linear costs, which capture queuing networks in cache networks. \notes{Extending \cite{li2020universally,li2021cache,mahdian2020kelly} our model could also capture queuing delay, yielding however a more complex objective than the one we encounter here.}
Wang et al. \cite{wang2016fair} analyze the proportional fairness of the total cost, Avrachenkov et al. \cite{avrachenkov2019distributed} study the fair caching problem in a video-on-demand system, and Liu et al. \cite{liu2021fair} consider fairness w.r.t. the utilities of caching gain
rates. Our model, objective, and, most importantly, constraints, significantly depart from the ones considered in the above works.

\noindent\textbf{Joint Optimization.}
Dehgan et al. \cite{dehghan2015complexity}, Poularakis et al. \cite{poularakis2020service, poularakis2013approximation}, Ioannidis and Yeh \cite{ioannidis2018jointly} and Liu et al. \cite{liu2019joint} consider the joint optimization of caching and routing in networks; the first two in particular study routing in the bipartite setting, while the last two do so in arbitrary topologies. Caching and routing decisions are formulated as binary variables in those works. Li et al. \cite{li2020universally,li2021cache} consider queuing networks and jointly optimize caching and service rate, which is a mixed integer optimization problem. Zafari et al. \cite{zafari2018optimal} jointly optimize data compression rate and data placement in a tree topology, posing this as a mixed integer problem; they solve this by a spatial branch-and-bound search strategy, which comes with no poly-time approximation guarantees.

Kamran et al. \cite{kamran2021rate} jointly optimize content placements and rate admission controls to avoid congestion in the cache network. Mentioning congestion control,  Dehgan et al. \cite{dehghan2015complexity}, Poularakis et al. \cite{poularakis2020service, poularakis2013approximation} all consider link capacities constraints. However, their network model and results do not apply to arbitrary topologies. 

Closer to us, Liu et al.~\cite{liu2019joint} consider  arbitrary topologies, and provide approximation guarantees, but have a different objective (throughput maximization) and constraints. In particular, there is no notion of routing costs, as incorporated in our setting. Moreover, their problem setup,  objective, and constraints, do not give rise to the DR-submodular structures we observe in our problem; altogether, their proposed algorithms cannot be applied to solve the optimization problem we encounter in our setting.

\noindent\textbf{Submodular Maximization.} Maximizing a monotone submodular function subject to a matroid constraint is classic. Krause and Golovin \cite{krause2014submodular} show that the greedy algorithm achieves a $1/2$ approximation ratio. Calinescu et al. \cite{calinescu2011maximizing} propose a \emph{continuous greedy} algorithm improving the ratio to $1-1/e$, that applies a Frank-Wolfe \cite{bertsekas1999nonlinear} variant to the multilinear extension of the submodular objective. With the help of auxiliary potential functions, Filmus and Ward \cite{filmus2014monotone} run a non-oblivious local search after the greedy algorithm, and also produce a $1-1/e$ approximation ratio. Further improvements are made by Sviridenko et al. \cite{sviridenko2017optimal} for a more restricted class of submodular functions with bounded curvature. Bian et al. \cite{bian2017guaranteed} show that the same Frank-Wolfe variant can be used to maximize continuous DR-submodular functions within a $1-1/e$ ratio. 

In  followup work, Bian et al. \cite{bian2017continuous} also show that another Frank-Wolfe variant achieves $1/e$ approximation guarantee for non-monotone DR-submodular function. Nevertheless, all these algorithms require either matroid constraints for set functions, or down-closed convex constraint for continuous functions. For general convex constraints, Hassani et al. \cite{hassani2017gradient} prove that projected gradient ascent yields a 1/2 approximation factor from the optimal. More close to our setting, Iyer et al. \cite{iyer2013submodular} and Crawford et al. \cite{crawford2019submodular} minimize/maximize a submodular function subject to a submodular/supermodular function inequality. However, none of above solve submodular maximization problems under multiple supermodular constraints in the continuous domain; this is the structure of the problems we consider here.

}{}

\section{Model}
\label{sec:model}
\arxiv{\begin{figure}[t]
    \centering
    \includegraphics[width = 0.9\linewidth]{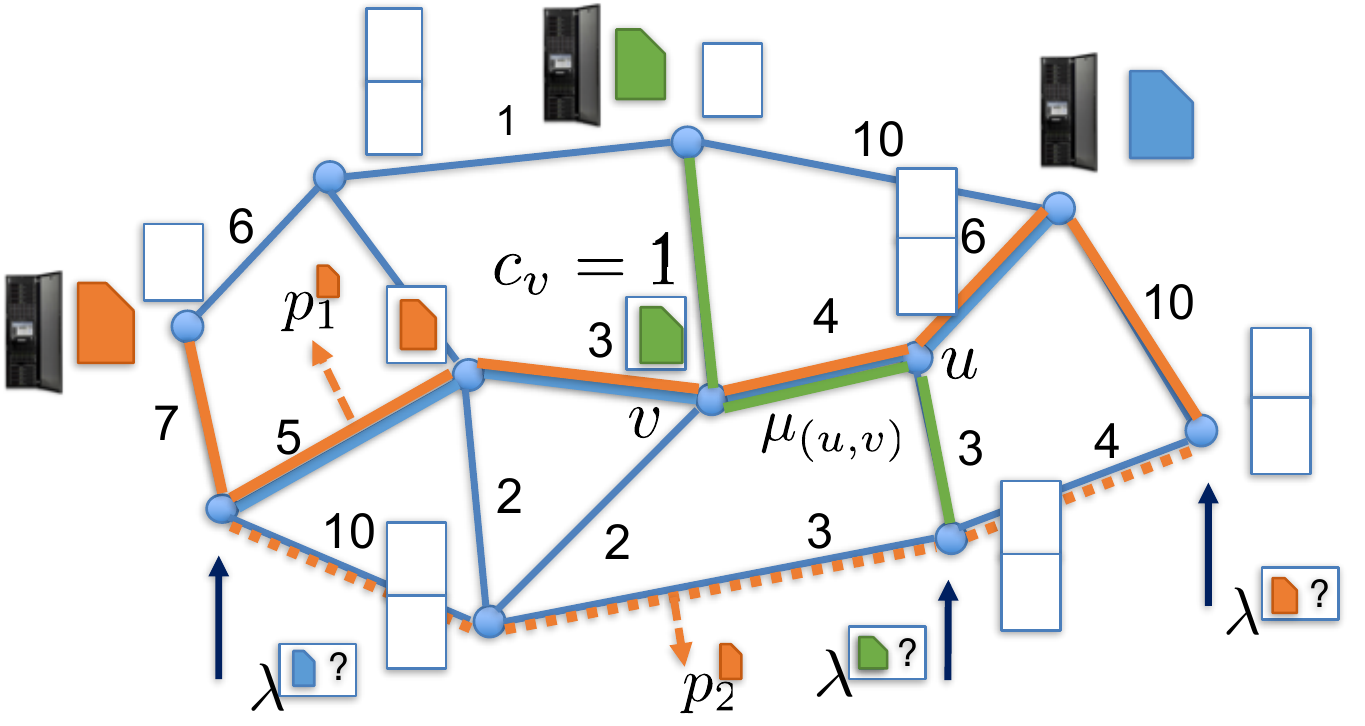}
    \caption{A cache network.}
    \label{fig:network}
\end{figure}}{}
\arxiv{We follow the model of Ioannidis and Yeh~\cite{ioannidis2018jointly}, with (a) an additional constraint on network capacities and (b) a shifted focus on probabilistic strategies w.r.t.~both caching and routing. We also depart by considering a more general request arrival process, rather than Poisson. Our model is illustrated in Fig.~\ref{fig:network}.}{}

\noindent\textbf{Network Model and Content Requests.}
We consider a network represented as a directed, symmetric\footnote{A directed graph is symmetric when $(i,j)\in E$ implies that $(j,i)\in E$.} graph $G(V,E)$.  
Content items (e.g., files, or file chunks) of equal size\arxiv{\footnote{This is w.l.o.g; see Section~\ref{sec:requestarrivals} below.}}{} are to be distributed across network nodes.  We denote by  $\catalog$ the set of  content items, i.e., the \emph{catalog}.
The network serves requests for items in $\catalog$ routed over the $G$.  A request $(i,s)$ is determined by (a) the  item $i\in \catalog$ requested, and (b) the request source $s\in V$. 
 We denote by $\requests\subseteq \catalog\times V$  the set  of all requests. 
\arxiv{As in prior work \cite{ioannidis2018jointly,ioannidis2016adaptive}, for}{For} each  $i\in \catalog$, there exists a fixed set of \emph{designated server} nodes $\source_i\subseteq V$, that always store $i$. A node $v\in \source_i$ permanently stores $i$ in \emph{excess memory outside its cache}. \arxiv{Thus, the placement of items to designated servers is fixed and outside the network's design.}{}
A request $(i,s)$ is routed over a path in $G$ towards a designated server. However, forwarding terminates upon reaching \emph{any intermediate cache} that stores $i$. At that point, a response carrying $i$ is sent over the reverse path,\arxiv{ i.e., from the node where the cache hit occurred,}{} back to  $s$. Both caching \emph{and} routing decisions are network design parameters, while request arrivals are problem inputs. We define all three below.

\noindent\textbf{Request Arrival Process. } \label{sec:requestarrivals} Requests  arrive according to an i.i.d.~process: time is slotted and, at each timeslot $t\in\naturals$, a random subset $\requests(t) \subseteq \requests$ of requests occur. \arxiv{We make no prior assumption on the distribution of i.i.d.~variables $\requests(t)$.}{}%
We denote by
\begin{align}
 \arriveprob_{(i,s)}=\prob[(i,s)\in \requests(t)]  \in [0,1], \quad (i,s)\in \requests,
\end{align}
the marginal probability that request $(i,s)$ occurs.
 
\arxiv{We note that the assumption that items are of equal size comes without any loss of generality \notes{\cite{ioannidis2016adaptive,ioannidis2018jointly,li2020universally}}. This is precisely because a set of requests $\requests(t) \subseteq \requests$ are issued at each timeslot. Hence, files of unequal size, split in equal-size chunks, can be captured in our model through  simultaneous requests of all of their constituent chunks from the same source $s$ within a single slot. We also note that, for notational simplicity, we assume that $\requests(t)$ is a set (i.e., each element $(i,s)\in\requests$ appears at most once), but our analysis can be easily extended to the case where $\requests(t)$ is a multiset (i.e., an item $i$ is requested multiple times from the same node $s$). This would be the case if, e.g., a node represents an access point serving multiple end users, that issue overlapping requests.}{}

\arxiv{
\renewcommand{\arraystretch}{1.1}
\begin{table}[!t]
\caption{Notation Summary}
\begin{tabular}{p{0.12\linewidth}p{0.75\linewidth}}
\hline
\multicolumn{2}{c}{\textbf{Common Notation}}\\
\hline
$G(V,E)$ & Network graph, with nodes $V$ and edges $E$\\
$\catalog$ & Item catalog\\
$c_v$ & Cache capacity at node $v\in V$\\
$\requests$ & Set of requests $(i,s)$, with $i\in\catalog$ and source $s\in V$\\
$\lambda_{(i,s)}$ & Marginal probability that request $(i,s)\in \requests$ \\
$\source_i$ & Set of designated servers of $i\in \catalog$\\
$x_{v,i}$ & Variable indicating whether $v\in V$ stores $i\in\catalog$ \\
$\cacheprobvec_{v,i}$ & Marginal probability that $v$ stores $i$\\ 
$X$ & Global caching strategy  of $x_{v,i}$s, in $\{0,1\}^{|V|\times |\catalog|}$\\
$\cacheprobvec$ & Expectation of caching strategy matrix $X$\\
$w_{u,v}$ & weight/cost of edge $(u,v)$\\

\hline
\multicolumn{2}{c}{\textbf{Source Routing}}\\
\hline
$\pathset_{(i,s)}$ & Set of paths request $(i,s)\in \requests$ can follow\\
$P_\SR$ & Total number of paths\\
$p$ & A simple path of $G$\\
$k_{p}(v)$ & The position of node $v\in p$ in path $p$.\\ 
$r_{(i,s),p}$ & Variable indicating  whether $(i,s)\in\requests$ is forwarded over $p\in\pathset_{(i,s)}$ \\
$\routeprob_{(i,s),p}$ & Marginal probability that $s$ routes request for $i$ over $p$\\ 
$r$ & Routing strategy of $r_{(i,s),p}$s,  in $\{0,1\}^{\sum_{(i,s)\in \requests}|\pathset_{(i,s)}|}$.\\
$\routeprobvec$ & Expectation of routing strategy vector $r$\\
\hline
\end{tabular}
\end{table}
}{}

\noindent\textbf{Caching Strategies.}
Each node has a cache that can store a finite number of items. We denote by $\capacity_v\in \naturals$ the  capacity at node $v\in V$\arxiv{: exactly $\capacity_v$ content items can be stored in $v$}{}.
For each node $v\in V$, vector $\cachestatevec_v\in\{0,1\}^{|\catalog|}$  indicates $v$'s caching state: $\cachestate_{v,i} \in \{0,1\},$  for $ i \in \catalog $,
is the binary variable indicating whether $v$ stores content item $i$. We assume that vectors $\cachestatevec_v$ are random and independent  across $v\in V$. As $v$ can store no more than $c_v$ items, we have: 
\begin{align}
\sum_{i\in \catalog} \cachestate_{v,i} \leq c_v,\text{ for all }v\in V.\label{capconst}
\end{align}
\arxiv{The  global caching state is the vector
 $\cachestatevec=[\cachestate_{v,i}]_{v\in V,i \in \catalog}\in \{0,1\}^{|V| |\catalog|},$ 
whose elements comprise the node caching state variables. }{}

We define the system's \emph{caching strategy} to be a stationary probability distribution over valid caching states $\cachestatevec\in \{0,1\}^{|V||\catalog|}$, i.e., ones that (a) satisfy Eq.~\eqref{capconst} and (b) have a product form over $v\in V$\arxiv{ (as states $\cachestatevec_v$, $v\in V$, are independent)}{}.
We denote by
\begin{align}
\cacheprob_{v,i} \equiv \prob[\cachestate_{v,i} = 1] = \expect[\cachestate_{v,i}]\in [0,1], \text{ for }i\in \catalog,
\label{cacheprob}
\end{align}
the marginal probability that node $v$ caches item $i$, and by 
\arxiv{$\cacheprobvec=[\cacheprob_{v,i}]_{v\in V,i\in \catalog}=\expect[\cachestatevec]\in [0,1]^{|V| |\catalog|},$
}{
$\cacheprobvec=[\cacheprob_{v,i}]_{v\in V,i\in \catalog}\in [0,1]^{|V| |\catalog|},$
}
the corresponding expectation of the caching strategy. By Eq.~\eqref{capconst} and Eq.~\eqref{cacheprob}:
\begin{align}
\label{cons: cache}
    \sum_{i\in \catalog} \cacheprob_{v,i}\leq c_v, \text{for all}~v\in V.
\end{align}

\noindent\textbf{Source Routing Strategies. }
\arxiv{Recall that requests are routed towards designated server nodes. }{}For every request $(i,s)\in \requests$, we assume that there exists a set $\pathset_{(i,s)}$ of \emph{paths} that the request can follow towards a designated server in $\source_i$. A source node $s$ can forward a request among any of these paths; however,  responses are constrained to   reversely follow the same path as the request  they  serve.
%
%
A path $\ppath$ of length $|p|=K$ is a sequence $\{p_1,p_2,\ldots,p_K\}$ of nodes $p_k\in V$\arxiv{ such that  $(p_k,p_{k+1})\in E$, for every $k\in \{1,\ldots,|p|-1\}$}{}.  
\arxiv{Following \cite{ioannidis2018jointly,ioannidis2016adaptive}, we assume that  paths in $\pathset_{(i,s)}$ are \emph{well-routed}, i.e.,   they satisfy the  four natural conditions:  for every $\ppath\in\pathset_{(i,s)}$:
(a) $p$ starts at $s$, i.e., $p_1=s$;
(b) $p$  is simple, i.e., it contains no loops;
(c) the last node in $p$  is a designated server for item $i$, i.e., if $|p|=K$, 
$p_K\in \source_i$; and
(d) no other node in $p$ is a designated server for $i$, i.e., if $|p|=K$,
$p_k\notin \source_i$, for $k=1,\ldots,K-1.$ }{}
Given a path $p$ and a $v\in p$, let   $k_p(v)$ be the position of $v$ in $p$\arxiv{; i.e., $k_p(v)$ equals to $k\in \{1,\ldots,|p|\}$ such that $p_k=v$}{}. 
Given sets $\pathset_{(i,s)}$, $(i,s)\in \requests$, the \emph{routing state} of a source $s\in V$ w.r.t.~request $(i,s)\in \requests$ is a vector $\routevec_{(i,s)}\in \{0,1\}^{|\pathset_{(i,s)}|}$, where
 $\route_{(i,s),p}\in \{0,1\}$   is a binary variable indicating whether $s$ selects path $p \in \pathset_{(i,s)}$. 
These satisfy:  
\begin{align}
\sum_{p\in \pathset_{(i,s)}} \route_{(i,s),p} =1,\text{  for all }(i,s)\in\requests,
\label{routecap}
\end{align}
indicating that exactly one path is selected. We again assume that $\routevec_{(i,s)}$ are independent random variables across $(i,s)\in \requests$.\arxiv{ Let
$P_\SR =\sum_{(i,s)\in \requests}|\pathset_{(i,s)}|,$
be the total number of paths.
We refer to the vector 
$
    \routevec= [ \routevec_{(i,s),p}]_{(i,s)\in \requests,p\in \pathset_{(i,s)}}\in  \{0,1\}^{P_\SR}
$
as the  global routing state vector. }{}

\sloppy
The system's \emph{routing strategy} to be a stationary distribution over valid routing states, i.e., states that (a) satisfy Eq.~\eqref{routecap} and (b) have a product form over $(i,s)\in \requests$\arxiv{ (as routing states $\routevec_{(i,s)}$ are independent across $(i,s)\in \requests$)}{}.  For $p\in\pathset_{(i,s)}$, let
\begin{align}
\routeprob_{(i,s),p} \equiv \prob[\route_{(i,s),p}=1 ] = \expect[\route_{(i,s),p}] \in [0,1]
\label{marginalroute},
\end{align}
be the marginal probability that path $p$ is selected by $s$. \arxiv{}{Let
$P_\SR =\sum_{(i,s)\in \requests}|\pathset_{(i,s)}|,$
be the total number of paths. }Then, the routing strategy is  determined by
\arxiv{$
\routeprobvec = [\routeprob_{(i,s),p}] \allowbreak _{(i,s)\in\requests ,p\in \pathset_{(i,s)} } = \expect[\routevec]\in [0,1]^{P_\SR},
$}{
$\routeprobvec = [\routeprob_{(i,s),p}] \allowbreak _{(i,s)\in\requests,p\in \pathset_{(i,s)} }   \in [0,1]^{P_\SR},$
}%
where, by Eqs.~\eqref{routecap} and \eqref{marginalroute},
\begin{align}
\label{cons: route}
\sum_{p\in \pathset_{(i,s)}} \routeprob_{(i,s),p} =1,~\text{for all}~(i,s)\in \requests.
\end{align}

\fussy

\noindent\textbf{Link Capacities. }
Every edge $(u,v)\in E$ is associated with a capacity $\mu_{u,v}\geq 0$, indicating  the maximum  traffic it can sustain: in expectation, the  traffic  at $(u,v)$ must not exceed $\mu_{u,v}$. Formally, since cache states across nodes in the path $p\in \pathset_{(i,s)}$ are independent, we have that for all $(u,v)\in E$:
\begin{align}
\label{cons: link_capacity}
\sum_{(i,s)\in \requests}\arriveprob_{(i,s)} \!\! \sum_{p\in \pathset_{(i,s)}:(v,u)\in p} \!\!\! \routeprob_{(i,s),p} \prod_{k'=1}^{k_p(v)} (1\!-\!\cacheprob_{p_{k'},i})  \leq \mu_{u,v}.
\end{align}
\noindent\textbf{Costs and Objective. }
To capture  costs  (e.g., latency, money, etc.), we  associate a \emph{weight} $w_{u,v}\geq 0$ with each  edge $(u,v)\in E$, representing the cost of transferring an item across $(u,v)$.  
\arxiv{We assume  that costs are solely due to response messages that carry an item, while request forwarding costs are negligible.}{} 
\arxiv{ We assume that costs are non-symmetric, i.e., $w_{u,v}\neq w_{v,u}$, generally.}{} 
Again, by independence,  the expected transfer cost for serving a request $(i,s)\in \requests$   given pair $(\cacheprobvec,\routeprobvec)$  is: 
\begin{align} 
C_{(i,s)} (\cacheprobvec,\routeprobvec) =\!\! \sum_{p\in \pathset_{(i,s)}} \!\! \routeprob_{(i,s),p} \!\sum_{k=1}^{|p|-1}\! w_{p_{k+1},p_k}\!\prod_{k'=1}^k (1\!-\!\cacheprob_{p_{k'},i}).\label{sourcecost}
\end{align}
Intuitively, Eq.~\eqref{sourcecost} states that $C_{(i,s)}$ includes the cost of an edge $(p_{k+1},p_k)$ in the path $p$ if (a) $p$ is selected by the routing strategy, and (b) \emph{no} cache preceding this edge in $p$  stores $i$.  

We wish to minimize the total expected transfer cost: 
\begin{subequations}
\label{relaxedmin}
\arxiv{\center{\textsc{MinCost}}\vspace*{-0.5em}}{}
\begin{align} 
\text{Minimize:} &\quad  C(\cacheprobvec,\routeprobvec) = \sum_{(i,s)\in \requests}\arriveprob_{(i,s)} C_{(i,s)} (\cacheprobvec,\routeprobvec) \\
\text{subj.~to:} &\quad\text{Eqs.}~\eqref{cacheprob}, \eqref{cons: cache}, \eqref{marginalroute}, \eqref{cons: route}, \text{ and }\eqref{cons: link_capacity}.
\end{align}
\end{subequations}
This problem is \emph{NP-hard} \cite{shanmugam2013femtocaching, ioannidis2018jointly}. We note that, the constraint set is not a convex polytope, due to Eq.~\eqref{cons: link_capacity}, and the objective is not convex. \final{Compared to the setting considered by Ioannidis and Yeh~\cite{ioannidis2018jointly}, we account for additional capacity constraints via Eq.~\eqref{cons: link_capacity}, which in turn lead to the non-convexity of the constraint set.}

\section{Main Results}
\label{sec:main results}
Despite the lack of convexity of Problem \eqref{relaxedmin}, we show that after an appropriate change of variables the objective can be written as a continuous DR-submodular function~\cite{bian2017guaranteed}. This gives rise to a primal-dual heuristic, in which primal steps are approximable via a polytime algorithm.

\subsection{Conversion to a Continuous DR-submodular Problem}

To convert Problem~\eqref{relaxedmin} to a problem amenable through a solution via algorithms that exploit DR-submodularity, we first introduce the auxiliary variables, for all $p\in \pathset_{(i,s)},(i,s)\in\mathcal{R}$:
\begin{align}
\label{cons: auxiliary route}
   \tilde{\routeprob}_{(i,s),p} = 1- \routeprob_{(i,s),p} \in [0,1].
\end{align}
I.e., these are the ``complements'' of the routing variables; we also denote the corresponding vector comprising these complement variables by $\tilde{\routeprobvec}\in [0,1]^{P_\SR}.$ 
Let $C_0 \equiv \sum_{(i,s)\in \requests} \allowbreak \lambda_{(i,s)}\sum_{p\in \pathset_{(i,s)}}\sum_{k=1}^{|p|-1} w_{p_{k+1},p_k}.$
Observe that this is a universal constant, not depending or $\routeprobvec$ or $\cacheprobvec$. We define the objective:
\begin{align}
\label{eq:cache_gain}
\begin{split}
F(\cacheprobvec,\tilde{\routeprobvec}) &= C_0-C(\cacheprobvec,1-\tilde{\routeprobvec})\\
& =\sum_{(i,s)\in \requests}\lambda_{(i,s)} \sum_{p\in \pathset_{(i,s)}} \sum_{k=1}^{|p|-1}w_{p_{k+1},p_k}\cdot(1-\\
& \hspace{50pt} (1-\tilde{\routeprob}_{(i,s),p})\prod_{k'=1}^k (1-\cacheprob_{p_{k'}i})),
\end{split}
\end{align}
as the expected \emph{cache gain}. Observe that $F$ is monotone increasing w.r.t.~all of its variables. Thus, Prob.~\eqref{relaxedmin} is equivalent to the following cache gain maximization problem:
\begin{subequations}
\label{maxgain}
\begin{align} \text{Maximize: } & F(\cacheprobvec,\tilde{\routeprobvec})\displaybreak[0] \\
\text{subj.~to: } & \eqref{cacheprob}, \eqref{cons: cache}, \eqref{cons: auxiliary route}\\
\sum_{p\in \pathset_{(i,s)}} & (1-\tilde{\routeprob}_{(i,s),p}) =1, \text{ for all }(i,s)\in\requests,\displaybreak[0] \label{pselect_1}\\
G_{u,v} & (\cacheprobvec,\tilde{\routeprobvec}) \leq 0 ,\hspace{10pt} \text{ for all }(u,v)\in E,  \displaybreak[0] \label{cons: link_capacity2}
\end{align}
\end{subequations}
where we define the \emph{flow} over edge $(u,v)\in E$ to be
\begin{align}
\label{eq:flow}
\lambda_{(u,v)}(\cacheprobvec,\tilde{\routeprobvec}) = \!\!\sum_{(i,s)\in \requests} \sum_{\overset {p\in \pathset_{(i,s)}:}  {(v,u)\in p}} \!\!\! \lambda_{(i,s)} (1\!-\!\tilde{\routeprob}_{(i,s),p}) \! \prod_{k'=1}^{k_p(v)} (1\!-\!
\cacheprob_{p_{k'},i}),
\end{align}
and the \emph{overflow} at $(u,v)\in E$ to be
\begin{align}
\label{eq:overflow}
\begin{split}
G_{u,v}(\cacheprobvec,\tilde{\routeprobvec}) = \lambda_{(u,v)}(\cacheprobvec,\tilde{\routeprobvec})- \mu_{u,v}.
\end{split}
\end{align}
The objective is not concave, and the constraints involving overflow functions above are not convex. Nevertheless, the following can be shown using the earlier analysis of \cite{bian2017guaranteed, ioannidis2018jointly}:
\begin{lemma}\label{lem:drsub}
Function $F$, defined in Eq.~\eqref{eq:cache_gain}, is non-decreasing and continuous \final{diminishing-returns} (DR)-submodular,  and functions $G_{u,v}$, for all $(u,v)\in E$, defined in Eq.~\eqref{eq:overflow}, are non-increasing and continuous DR-supermodular.
\end{lemma}
\arxiv{Formally, a twice-differentiable function $F$ is continuous diminishing returns (DR) submodular (supermodular) \cite{krause2014submodular} if the off-diagonal elements of its Hessian $\nabla^2 F $ are non-positive (non-negative). }{}Existing algorithms for DR-submodular maximization \cite{bian2017guaranteed, hassani2017gradient, iyer2013submodular} do not directly apply to our optimization problem, as they require constraints either being convex or containing at most one supermodular constraint. \arxiv{Nevertheless, we exploit this property  in our primal-dual algorithm.}{}
\subsection{Lagrangian and Duality}
Consider the {Lagrangian}:
\begin{align}
\label{eq:lagrangian}
    L(\cacheprobvec,\tilde{\routeprobvec},\dualvec) =    F(\cacheprobvec,\tilde{\routeprobvec}) -\textstyle \sum_{e\in E} \dual_{u,v} \cdot G_{u,v}(\cacheprobvec,\tilde{\routeprobvec}) ,
\end{align}
where vector $\dualvec = [\dual_{u,v}]_{(u,v) \in E}$ is the non-negative \emph{dual variables} associated with the constraint \eqref{cons: link_capacity2}. Intuitively, the Lagrangian function $L$ penalizes the infeasibility of the link capacity constraints. The following theorem is an immediate consequence of Lemma~\ref{lem:drsub}:
\begin{theorem}\label{thm:drsub}
Function $L$ is non-decreasing and continuous DR-submodular.
\end{theorem}
To motivate our approach, assume we were given  proper dual variables $\dualvec$. Then, optimizing the Lagrangian converts the cache gain maximization problem \eqref{maxgain} to the following:
\begin{subequations}
\label{MaxSubmodular}
\begin{align} 
\text{Maximize:} &\quad  L(\cacheprobvec,\tilde{\routeprobvec},\dualvec)\\
\text{subj.~to:} &\quad \cacheprobvec,\tilde{\routeprobvec} \in \feasibledomain
\end{align}
\end{subequations}
where $\feasibledomain$ is the set defined by constraints: \eqref{cacheprob}, \eqref{cons: cache}, \eqref{cons: auxiliary route}, and \eqref{pselect_1}. Prob.~\eqref{MaxSubmodular} has a non-decreasing, continuous DR submodular objective, and convex constraints $\feasibledomain$. \arxiv{For arbitrary convex constraint,  projected gradient ascent~\cite{hassani2017gradient} achieves an $\frac{1}{2}$ approximate ratio. If, in addition, it were \emph{down-closed}, \footnote{A set $S\subset \reals^d_+$ is down closed if for all $\boldsymbol{x}\in S$ and all $\boldsymbol{x}'\in \reals^d_+$ for which $\boldsymbol{x}' \le \boldsymbol{x}$, $\boldsymbol{x}'\in S$.} a Frank-Wolfe algorithm variant~\cite{bian2017guaranteed,calinescu2011maximizing} would attain a more favorable $1-\frac{1}{e}$ approximation. Nevertheless, even though it is not down-closed, we show that this problem can indeed attain this improved approximation via a Frank-Wolfe algorithm (see Thm.~\ref{thm:FW variant}), by an appropriate relaxation of its constraints.}{}

\subsection{Primal-Dual Algorithm}
Motivated by the above observation, 
we propose solving Prob.~\eqref{maxgain} via a primal-dual algorithm. The primal steps of the algorithm reduce to solving, Prob.~\eqref{MaxSubmodular} which is a monontone DR-submodular optimization problem with affine constraints; though not down-closed convex or even not convex, we are able to solve this via a polytime algorithm within a $1-1/e$ approximation guarantee.

\subsubsection{Algorithm Overview}

\begin{algorithm}[t]
\caption{Primal-Dual Algorithm}
\label{alg:PrimalDual}
\LinesNumbered
\KwIn{$L(\jvar,\dualvec)$, $\feasibledomain$. }
$t \gets 0$, $\dualvec(0) \gets \boldsymbol{0}.$ \\
\While{$t< \tau$ \notes{convergence condition is not met}}{
 $\jvar(t+1) = \alpha_t \argmax_{\jvar\in \feasibledomain} L(\jvar,\dualvec(t)) + (1 - \alpha_t) \jvar(t)$ \\
 $\dual_e(t+1) = \left[ \dual_e(t) + \beta_t G_{e}\left(\jvar(t+1)\right) \right]^+, \text{ for all }e\in E$ \\
 $t \gets t+1$ 
}
\Return $\jvar_k$ \\
\end{algorithm}

For brevity, we join $\cacheprobvec(t)$ and $\tilde{\routeprobvec}(t)$ as one variable $\jvar(t) = (\cacheprobvec(t), \tilde{\routeprobvec}(t))$, and denote by it \emph{primal variables}. The primal-dual algorithm starts from $\dualvec(0)=\boldsymbol{0}$ and iterates over:
\begin{subequations}
\label{eq:pdalgo}
\begin{align}
    & \jvar(t+1) = \alpha_t \argmax_{\jvar\in \feasibledomain} L(\jvar,\dualvec(t)) + (1 - \alpha_t) \jvar(t), \label{eq: primal} \\
    & \dual_e(t+1) = \left[ \dual_e(t) + \beta_t G_{e}(\jvar(t\!+\!1) \right]^+, \text{for all }e\in E, \label{eq: dual}
\end{align}
\end{subequations}
where $\alpha_t = \frac{2}{t+2}$ is the parameter of momentum, $\beta_t = \frac{c}{\sqrt{t}}$ is the step size, $c$ is a constant, and $[z]^+ = \max \{z, 0\}$. We summarize this also in Alg.~\ref{alg:PrimalDual}, and discuss each step in detail below:

\noindent\textbf{Primal Step \eqref{eq: primal}}: The primal step updates primal variables $\jvar$ given dual variables $\dualvec$. It first solves Prob.~\eqref{MaxSubmodular}; then, it utilizes a momentum parameter to alleviate the change of primal variables. Since $\jvar(t+1)$ is a convex combination of two points in feasible set $\feasibledomain$, it still lies in $\feasibledomain$. We describe how to solve \eqref{MaxSubmodular} approximately in Sec.~\ref{sec:FrankWolfe}. The smoothing process via the momentum is crucial, as it helps with the convergence of the algorithm: we observe this experimentally in Sec. \ref{sec:convergence}.

\noindent\textbf{Dual Step \eqref{eq: dual}}: Finally, the dual step updates dual variables $\dualvec$ given primal variables $\jvar$ via dual ascent. 

\subsubsection{Primal Variables via Frank-Wolfe Algorithm}
\label{sec:FrankWolfe}
\begin{algorithm}[t]
\caption{Frank-Wolfe variant for $L(\jvar,\dualvec(t))$}
\label{alg:FWvariant}
\LinesNumbered
\KwIn{$L(\jvar,\dualvec(t))$, $\feasibledomain'$, step size $\gamma \in (0,1]$. }
$\tau \gets 0$, $k \gets 0$, $\jvar_0 \gets \boldsymbol{0}.$ \\
\While{$\tau<1$}{
 $\boldsymbol{v}_k \gets \mathop{\arg\max}_{\boldsymbol{v}\in \feasibledomain'} \langle \boldsymbol{v}, \nabla L(\jvar,\dualvec(t)) \rangle$ \\
 $\gamma_k \gets \min\{ \gamma, 1-\tau \}$ \\
 $\jvar_{k+1} = \jvar_k + \gamma_k\boldsymbol{v}_k$, $\tau\gets\tau+\gamma_k$, $k \gets k+1$ 
}
\Return $\jvar_k$ \\
\end{algorithm}

We solve Problem \eqref{eq: primal} through a variant of Frank-Wolfe algorithm, summarized in Alg.~\ref{alg:FWvariant}. Starting from  $\jvar_0 = (\cacheprobvec_0, \tilde{\routeprobvec}_0) =\boldsymbol{0}$, the \arxiv{variant of Frank-Wolfe }{}algorithm iterates over:
\begin{subequations}
\label{eq:FW}
\begin{align}
    &\boldsymbol{v}_k = \argmax_{\boldsymbol{v} \in \feasibledomain'} \langle \boldsymbol{v}, \nabla L(\jvar,\dualvec(t)) \rangle \\
    &\jvar_{k+1} = \jvar_{k} + \gamma_k\boldsymbol{v}_k,
\end{align}
\end{subequations}
where $\gamma_k$ is the proper step size satisfying $\sum_k \gamma_k = 1$, gradient
\begin{equation}
     \nabla_i L(\jvar,\dualvec(t)) = L(\jvar,\dualvec(t)|y_i\!=\!1)-L(\jvar,\dualvec(t)|y_i\!=\!0),
\end{equation} 
and $\feasibledomain'$ is the set:
\begin{subequations}
\label{eq:down closed convex}
\begin{align}  
& \eqref{cacheprob}, \eqref{cons: cache}, \eqref{cons: auxiliary route},\\
&\textstyle \sum_{p\in \pathset_{(i,s)}} (1-\tilde{\routeprob}_{(i,s),p}) \ge 1, \hspace{10pt} \text{ for all }(i,s)\in\requests, \label{pselect_2}
\end{align}
\end{subequations}
The difference between $\feasibledomain$ and $\feasibledomain'$  lies in having  inequalities in Eq.~\eqref{pselect_2}, which relaxes Eq.~\eqref{pselect_1}. Note that $\feasibledomain'$ is a down-closed convex set while $\feasibledomain$ is not. The following theorem states the approximation guarantee we attain for this algorithm \text{w.r.t.~the (non-relaxed) Prob.~\eqref{MaxSubmodular}.}

\begin{theorem}
\label{thm:FW variant}
Let $\jvar^*$ be an optimal solution to Prob.~\eqref{MaxSubmodular}, and $\jvar_{\texttt{FW}}$ be the output of the Frank-Wolfe variant  Alg.~\ref{alg:FWvariant}. Then, $\jvar_{\texttt{FW}}$ belongs to $\feasibledomain$, and given any $\dualvec$:
\begin{equation}
L(\jvar_{\text{FW}},\dualvec)+C \ge (1-\frac{1}{e}) (L(\jvar^*,\dualvec)+C) \notes{- \frac{M}{2K}},
\end{equation}
where constant $C = \sum_{e\in E} \dual_e (\sum_{(i,s)\in \requests} \lambda_{(i,s)}-\mu_e)$\notes{, $M = 2 L(\boldsymbol{1},\dualvec)(|V||\catalog|+P_\SR)^2$ is the Lipschitz continuous
constant, and $K = \frac{1}{\gamma}$ is the number of iterations.}
\end{theorem}
\arxiv{The proof can be found in Appendix \ref{sec:proof FW variant}.}{\noindent\final{\textbf{Proof Sketch.} Frank-Wolfe variant algorithm shown in Alg.~\ref{alg:FWvariant} is a classic method \cite{bian2017guaranteed} for:
\begin{equation}
    \max_{\jvar\in \feasibledomain' } L(\jvar,\dualvec),
\end{equation}
which is a continuous DR-submodular maximization problem under down-closed convex constraint.
We first prove that constraints $\feasibledomain'$ are binding, i.e., there exists an optimal point $y^{**} = \argmax_{\jvar\in \feasibledomain' } L(\jvar,\dualvec)$, such that the inequality \eqref{pselect_2} in $\feasibledomain'$ holds with equality \eqref{pselect_1} in $\feasibledomain$, hence, $\jvar^{**}\in\feasibledomain$. This part is proved by contradiction. Thus, we can infer that $L(\jvar^{**},\dualvec) = L(\jvar^{*},\dualvec)$.
Similarly, $\jvar_{\texttt{FW}} \in \feasibledomain$ is also in $\feasibledomain'$. Finally, to provide an optimality factor, we offset $L$ by a constant $C$. We provide a $1 - \frac{1}{e}$ performance guarantee according to Corollary 1 in \cite{bian2017guaranteed}.
$\hfill\qed$}}

\notes{Note that, if we choose a large enough $K$, the offset $\frac{M}{2K}$ can become arbitrary small.  The constant $C$ is necessary to obtain an approximation guarantee as, in general, the Lagrangian \eqref{eq:lagrangian} can become negative, and adding this term ensures positivity.  } 
\final{In practice, we found that setting the scaling factor $c$ in $\beta_t$, defined in \eqref{eq: dual}, so that the Lagrangian remains always positive is preferable experimentally: in some sense, ensuring the positivity of $L$ strikes a good balance between the two components (cache gain and constraint penalization) of the objective. In contrast, a negative Lagrangian indicates a high penalization of infeasibility, and a discount of the cache gain.}
%
\notes{Furthermore, given a gradient, algorithm \eqref{eq:FW} requires polynomial time in the number of constraints and variables, which are $O(|V||\mathcal{C}|+|E||\mathcal{R}|)$. We iterate \eqref{eq:FW} at most $O(|V||\mathcal{C}|)$ times \cite{mahdian2020kelly}.}

\section{Experiments}
\label{sec:experiments}
\begin{table*}[t]
\caption{\notes{Graph Topologies and Experiment Parameters}}
\centering
\begin{small}
\begin{tabular}{ccccccccccc}
Graph & $|V|$ & $|E|$ & $|\mathcal{Q}|$ & $|\requests|$ & $|\pathset_{(i,s)}|$ & $|\capacity_v|$ & $w$ & $|\catalog|$ & $F_{\texttt{PD}}^1$ & $F_{\texttt{PD}}^3$ \\
\hline
\multicolumn{11}{c}{synthetic topology experiments}\\
\hline
\arxiv{\texttt{ER}}{Erd\H{o}s-R\'enyi (\texttt{ER})} & 100 & 1044 & 10 & 4949 & 1-5 & 10-20 & 1-100 & 1000 & 2314.9 & 2318.1\\
\arxiv{\texttt{BT}}{balanced tree (\texttt{BT})} & 364 & 726 & 10 & 4988 & 1-5 & 10-20 & 1-100 & 1000 & 1665.2 & 1666.3 \\
\arxiv{\texttt{HC}}{hypercube (\texttt{HC})} & 128 & 896 & 10 & 4960 & 1-5 & 10-20 & 1-100 & 1000 & 3228.5 & 3229.4\\
\arxiv{\texttt{grid}}{grid\_2d (\texttt{grid})} & 100 & 360 & 10 & 4954 & 1-5 & 10-20 & 1-100 & 1000 & 5753.2 & 5753.7\\
\arxiv{\texttt{SW}}{small-world (\texttt{SW})~\cite{kleinberg2000small}} & 100 & 503 & 10 & 4953 & 1-5 & 10-20 & 1-100 & 1000 & 4482.1 & 4484.3\\
\texttt{Ex1}  & \multirow{2}{*}{7} & \multirow{2}{*}{14} & \multirow{2}{*}{2} & \multirow{2}{*}{3} & \multirow{2}{*}{1-2} & \multirow{2}{*}{0-1} & \multirow{2}{*}{1-100} & \multirow{2}{*}{2} & 398.8 & 388.7\\
\texttt{Ex2} & & & & & & & & & 351.8 & 365.4\\
\hline
\multicolumn{11}{c}{backbone network experiments \arxiv{}{\cite{rossi2011caching}}} \\
\hline
\texttt{GEANT} & 22 & 66 & 4  & 4761 & 1-5 & 10-20 & 1-100 & 1000 & 4436.2 & 4440.7\\ 
\arxiv{\texttt{DT}}{Deutsche Telekom (\texttt{DT})} & 68 & 546 & 4 & 4929
& 1-5 & 10-20 & 1-100 & 1000 & 2014.7 & 2030.0 \\
\texttt{Abilene1}  & \multirow{2}{*}{11} & \multirow{2}{*}{28} & \multirow{2}{*}{3} & \multirow{2}{*}{4} & \multirow{2}{*}{1-2} & \multirow{2}{*}{0-1} & \multirow{2}{*}{1-100} & \multirow{2}{*}{4} & 814.3 & 901.0 \\
\texttt{Abilene2} & & & & & & & & & 761.4 & 789.4  \\ 
\hline
\multicolumn{11}{c}{trace-driven experiments} \\
\hline
\texttt{KS1} & 152 & 22952 & 101 & 1988 & 1-5 & 25-3195 & 1-100 & 526 & 19938.5 & 19946.7\\
\texttt{KS2} & 152 & 22952 & 103 & 4963 & 1-5 & 50-6390 & 1-100 & 1207 & 35353.1 & 35349.4 
\end{tabular}
\end{small}
\label{tab:setting}
\end{table*}
\arxiv{
\begin{figure*}[t]
\centering
\begin{minipage}{0.65\linewidth}
    \centering
    \subcaptionbox{\texttt{Ex1} with $\kappa = 1$\label{fig:topology_example1}}{\includegraphics[width = 0.33\linewidth]{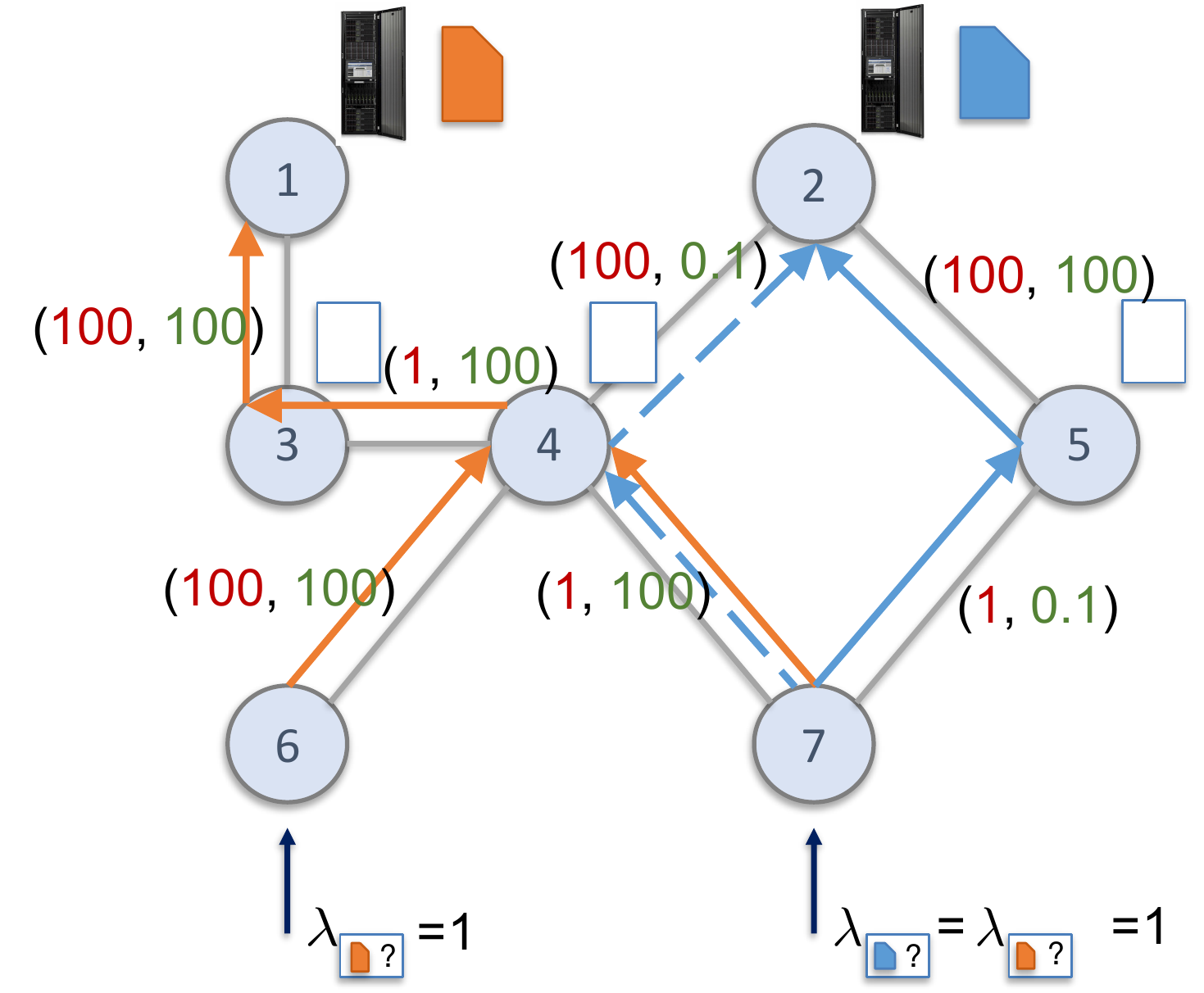}}
    \subcaptionbox{\texttt{Abilene1}  with $\kappa = 1$\label{fig:topology_abilene1}}{\includegraphics[width = 0.63\linewidth]{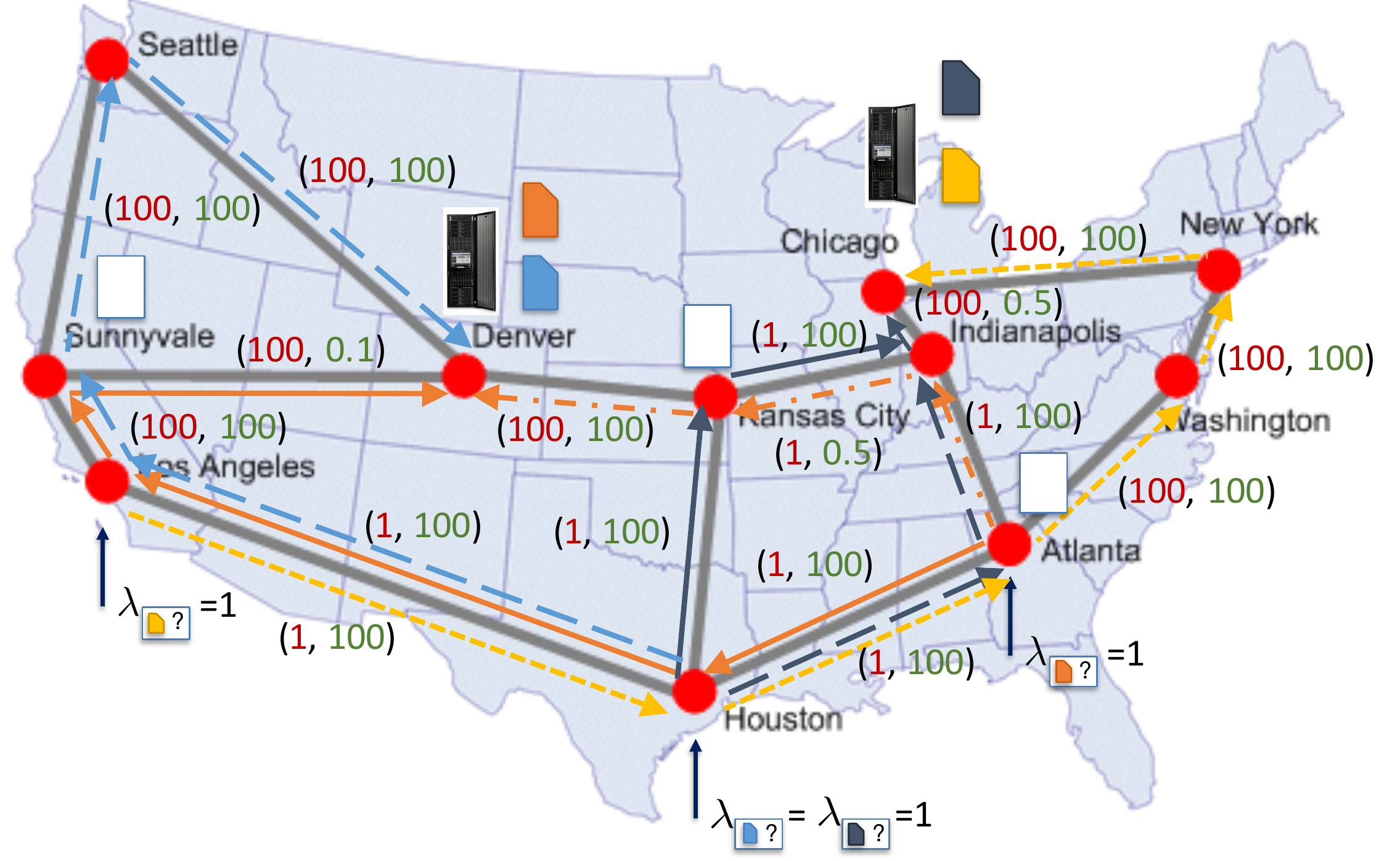}}
    \caption{Topologies and parameters of \texttt{Ex1} and \texttt{Abilene1} with designed requests and bandwidths. There is a pair (\textcolor{red}{\texttt{red}}, \textcolor[rgb]{0.1,0.6,0.3}{\texttt{green}}) for edge (u,v), where the first \textcolor{red}{\texttt{red}} number is the weight $w_{(v,u)}$ and the second \textcolor[rgb]{0.1,0.6,0.3}{\texttt{green}} number is the link capacity $\mu_{(v,u)}$.}
\end{minipage}
\hspace{5mm}
\begin{minipage}{0.25\linewidth}
    \centering
    \includegraphics[width = 1.0\linewidth]{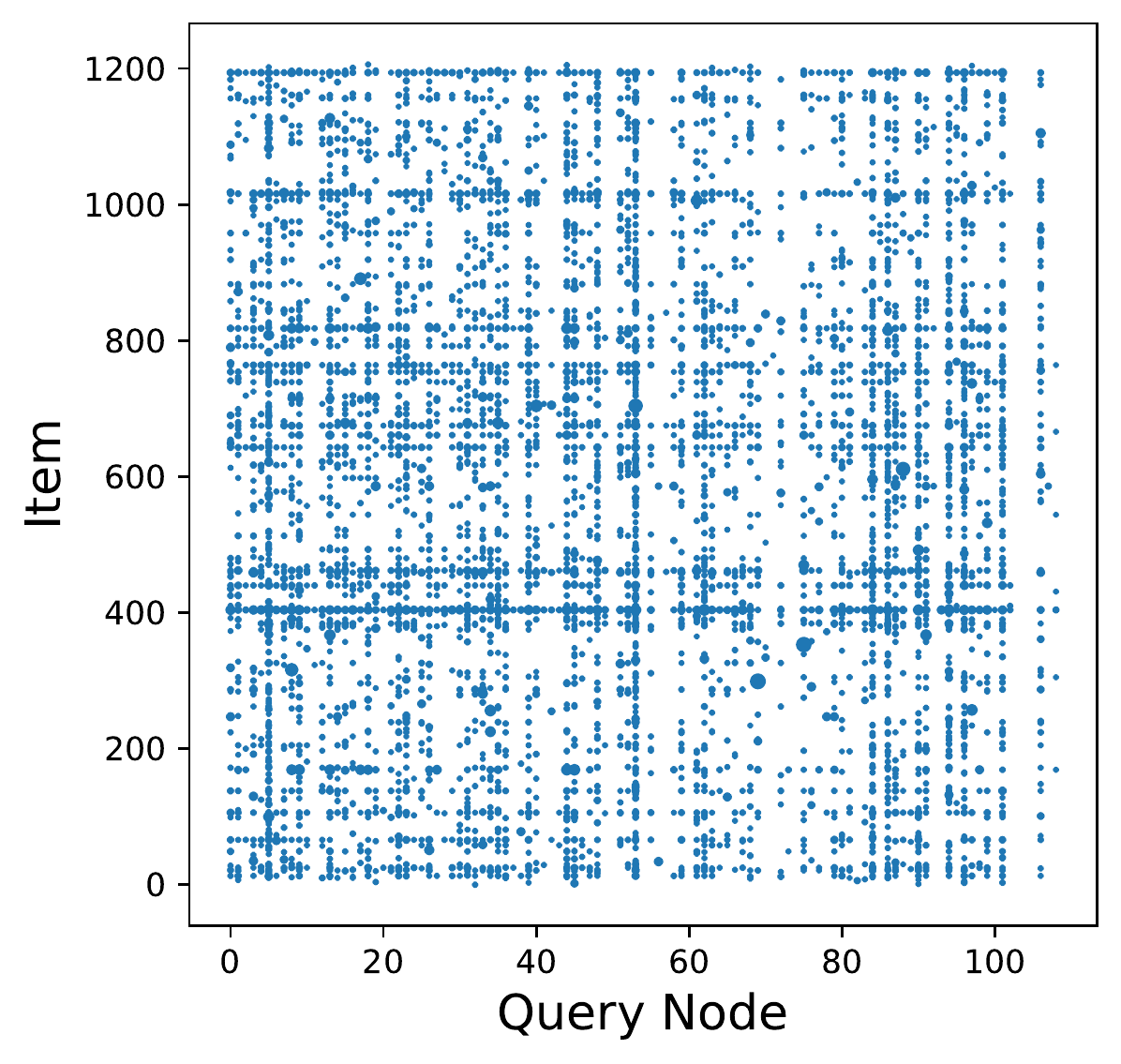}
    \caption{Request distribution for \texttt{KS2}. The radius of each point is proportional to the number of requests for that item during the observation period.}
    \label{fig:trace_real}
\end{minipage}
\end{figure*}
}{
\begin{figure}
    \centering
    \subcaptionbox{\texttt{Ex1} with $\kappa = 1$\label{fig:topology_example1}}{\includegraphics[width = 0.37\linewidth]{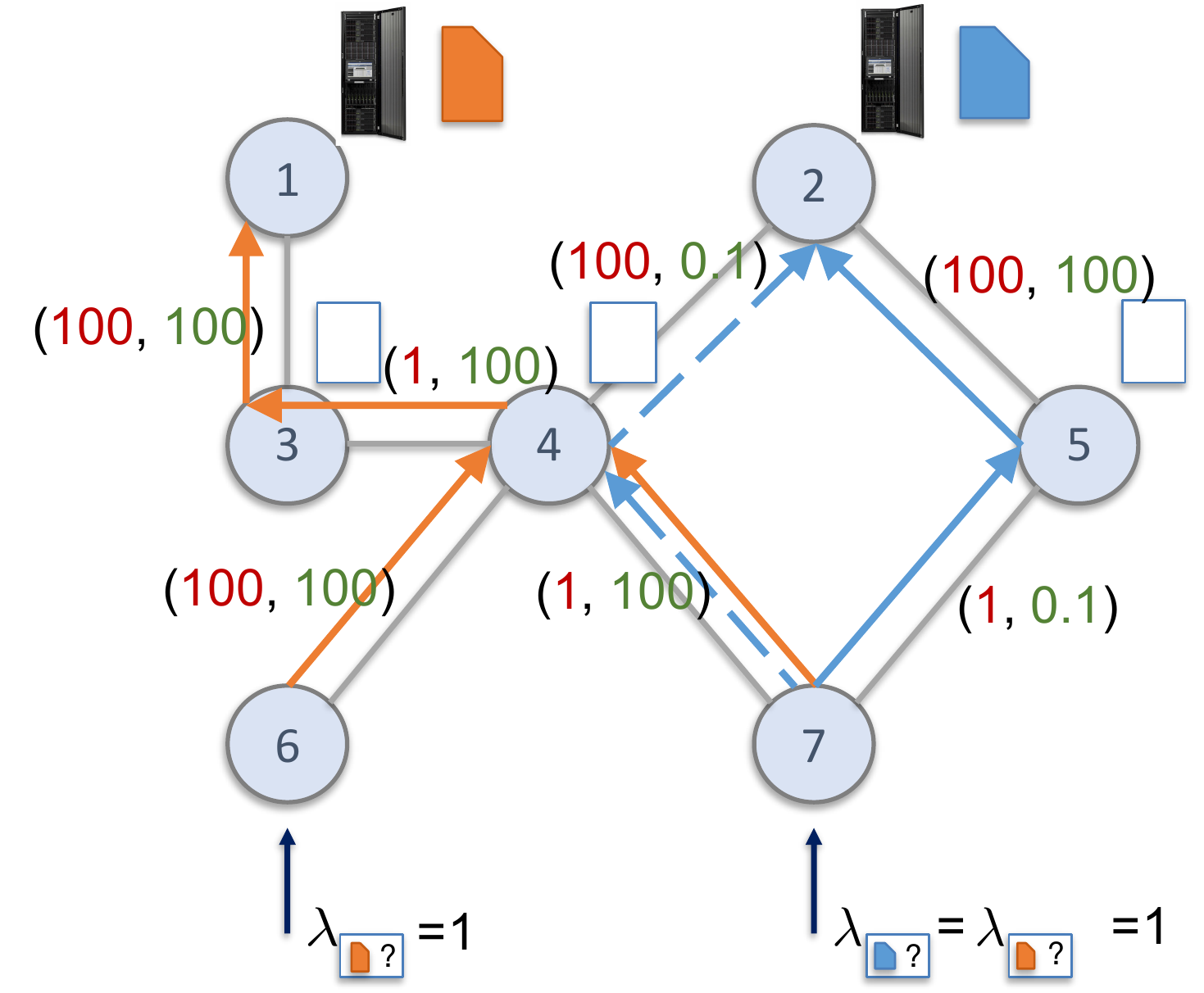}}
    \subcaptionbox{\texttt{Abilene1}  with $\kappa = 1$\label{fig:topology_abilene1}}{\includegraphics[width = 0.61\linewidth]{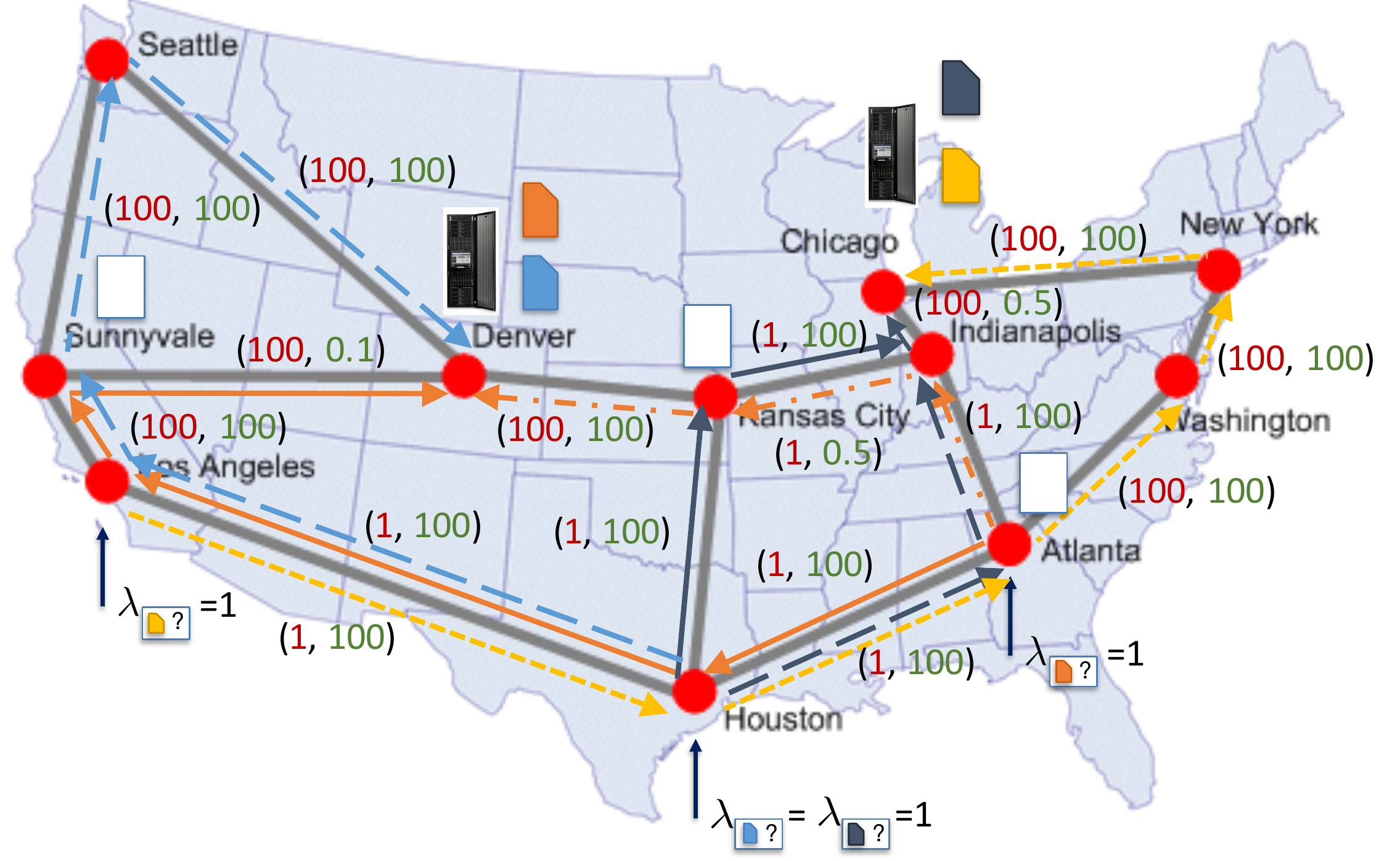}}
    \caption{Topologies and parameters of \texttt{Ex1} and \texttt{Abilene1} with designed requests and bandwidths. There is a pair (\textcolor{red}{\texttt{red}}, \textcolor[rgb]{0.1,0.6,0.3}{\texttt{green}}) for edge (u,v), where the first \textcolor{red}{\texttt{red}} number is the weight $w_{(v,u)}$ and the second \textcolor[rgb]{0.1,0.6,0.3}{\texttt{green}} number is the link capacity $\mu_{(v,u)}$.}
\end{figure}
}

We conduct both \emph{synthetic} and \emph{trace-driven} experiments.

\subsection{Synthetic Experiment Setup}
\sloppy
\noindent\textbf{Networks.} To evaluate our proposed algorithm, we perform experiments over five synthetic graphs\arxiv{, namely, Erd\H{o}s-R\'enyi (\texttt{ER}), balanced tree (\texttt{BT}), hypercube (\texttt{HC}), grid\_2d (\texttt{grid}), small-world (\texttt{SW})~\cite{kleinberg2000small}}{}, and a counter example designed to demonstrate suboptimality of competitors (\texttt{Ex}). We also experiment with three backbone network topologies\arxiv{: Deutsche Telekom (\texttt{DT}), \texttt{GEANT}, \texttt{Abilene} \cite{rossi2011caching}}{}. The parameters of different topologies are summarized in Tab.~\ref{tab:setting}. 
\final{The weights of each edge $w_{u,v}$, $(u,v)\in\mathcal{E}$ are selected uniformly at random (u.a.r.) from 1 to 100. Each node $v \in V$ has $\capacity_v$ storage to cache items from a catalog of size $|\catalog|$. Each item $i\in \catalog$ is stored permanently in one designated server $\mathcal{S}_i$ which is picked u.a.r. from $V$; the item is stored outside the designated server's cache. }
For \texttt{Ex} and \texttt{Abilene}, we select parameters in a way demonstrated in Figs. \ref{fig:topology_example1} and \ref{fig:topology_abilene1}, respectively. 

\fussy
\noindent\textbf{Requests. }
We generate requests synthetically as follows. 
\final{We select u.a.r.~a set of $\mathcal{Q}$ nodes from $V$ as the possible query nodes. The set of requests $\requests \subseteq \catalog \times \mathcal{Q}$ is then generated by sampling from the set $\catalog \times \mathcal{Q}$, u.a.r. }
For each such request $(i, s)\in \requests$, we select the request arrival probability $\arriveprob_{(i,s)}$ according to a Zipf distribution with parameter 1.2. 
\final{For each request $(i, s)\in \requests$, we generate at most $|\pathset_{(i,s)}|$ paths from the source $s\in V$ to the designated server $\mathcal{S}_i$, where the source $s$ and the designated server $\mathcal{S}_i$ are not the same node. In all cases, this path set includes the shortest path to the designated server. We consider only paths with stretch at most 4; that is, the maximum cost of a path in $\pathset_{(i,s)}$ is at most 4 times the cost of the shortest path to the designated source.} 
We follow a different synthetic request generation process for \texttt{Ex} and \texttt{Abilene}\arxiv{. Requests are designed}{,} based on the ``hard'' examples we describe in \arxiv{Appendix \ref{sec:proof suboptimal}}{\cite{li2022extended}}, on which we prove that competitors may fail to produce feasible solutions\arxiv{ (c.f.~Section~\ref{sec:Algorithms}). Parameter details are specified in Fig.~\ref{fig:topology_example1} and Fig.~\ref{fig:topology_abilene1}, for \texttt{Ex1} and \texttt{Abilene1}, respectively. Parameters for the remaining two topologies are described in Appendix \ref{sec:requests}}{.} 

\noindent\textbf{Link Capacities.} To control the level of congestion in the network, we determine link capacities $\mu_{u,v}, (u,v)\in E$ as follows.
We first \arxiv{assume random caching and routing, both set u.a.r. That is, we }{}randomly sample $\capacity_v$ items $i$ and set $\cacheprob_{v,i}=1$, for all $v\in V$, and set $\tilde{\routeprob}_{(i,s),p} = \frac{1}{|\pathset_{(i,s)}|}$, for all $p\in\pathset_{(i,s)}, (i,s)\in \requests$.
Then, we set the link capacities as $\mu_{u,v} = \kappa \lambda_{(u,v)}(\cacheprobvec,\tilde{\routeprobvec})$ correspondingly, where $\lambda_{(u,v)}$ is the flow on edge $(u,v)$, given by Eq.~\eqref{eq:flow}, and $\kappa \ge 1$ is a \emph{looseness coefficient}: the higher $\kappa$ is, the easier it is to satisfy the link capacity constraints. \arxiv{Note that, for every link $(u,v) \in E$, if $\mu_{u,v} \ge G'_{(u,v)}(\boldsymbol{0},\boldsymbol{0}) = \sum_{(i,s)\in \requests} \sum_{\overset {p\in \pathset_{(i,s)}:}  {(v,u)\in p}} \! \lambda_{(i,s)}$, then the link capacity constraint at that link is trivially satisfied. In our experiments, we set $1 \le \kappa < \max_{(i,s)\in \requests}|\pathset_{(i,s)}|$ to avoid this.
Link capacities of \texttt{Ex1} and \texttt{Abilene1} is given in Fig.~\ref{fig:topology_example1} and Fig.~\ref{fig:topology_abilene1} respectively.}{}

\subsection{Trace-Driven Experiment Setup}
Finally, we also conduct trace-driven simulations using data from a short video application, Kuaishou (\texttt{KS}) \cite{kuaishou}. This comprises more than 8 million requests of 2 million items/videos reaching 488 Kaishou edge servers deployed at 31 provinces in China \arxiv{from 8:00pm to 8:05pm on 12/04/2018}{within 5 mins}. The network topology (including nodes, links, and link and cache capacities) are determined from an actual cache deployment by Kuaishou. We preprocess the data to create two instances (\texttt{KS1} and \texttt{KS2}), whose statistics are summarized in Tab.~\ref{tab:setting}, as follows. 

\final{We select the largest connected subgraph, and utilize $\frac{1}{4}$ and $\frac{1}{2}$ of caches equipped by each node for our experiments \texttt{KS1} and \texttt{KS2}, respectively. In \texttt{KS1}, we restrict traffic of top 2000 popular requests, while in \texttt{KS2} we restrict traffic  to the top 5000 popular requests\arxiv{; the request distribution of latter is shown in Fig.~\ref{fig:trace_real}}{}. We again generate all paths of stretch at most 4; we drop any request that does not contain any paths in the largest connected component, leading to the numbers reported in Table~\ref{tab:setting}. We use these to compute request probabilities $\lambda_{(i,s)}\in[0,1]$; to do so, we normalize each request frequency by the frequency of the most popular request. As we limit traffic to a subset of the entire demand, we scale link capacities in \texttt{KS1} and \texttt{KS2} both by $\frac{1}{2250}$.}

\arxiv{\begin{figure*}[t]
    \centering
    \subcaptionbox{Looseness coefficient $\kappa=1$\label{fig:topologies1}}{\includegraphics[width = 0.9\linewidth]{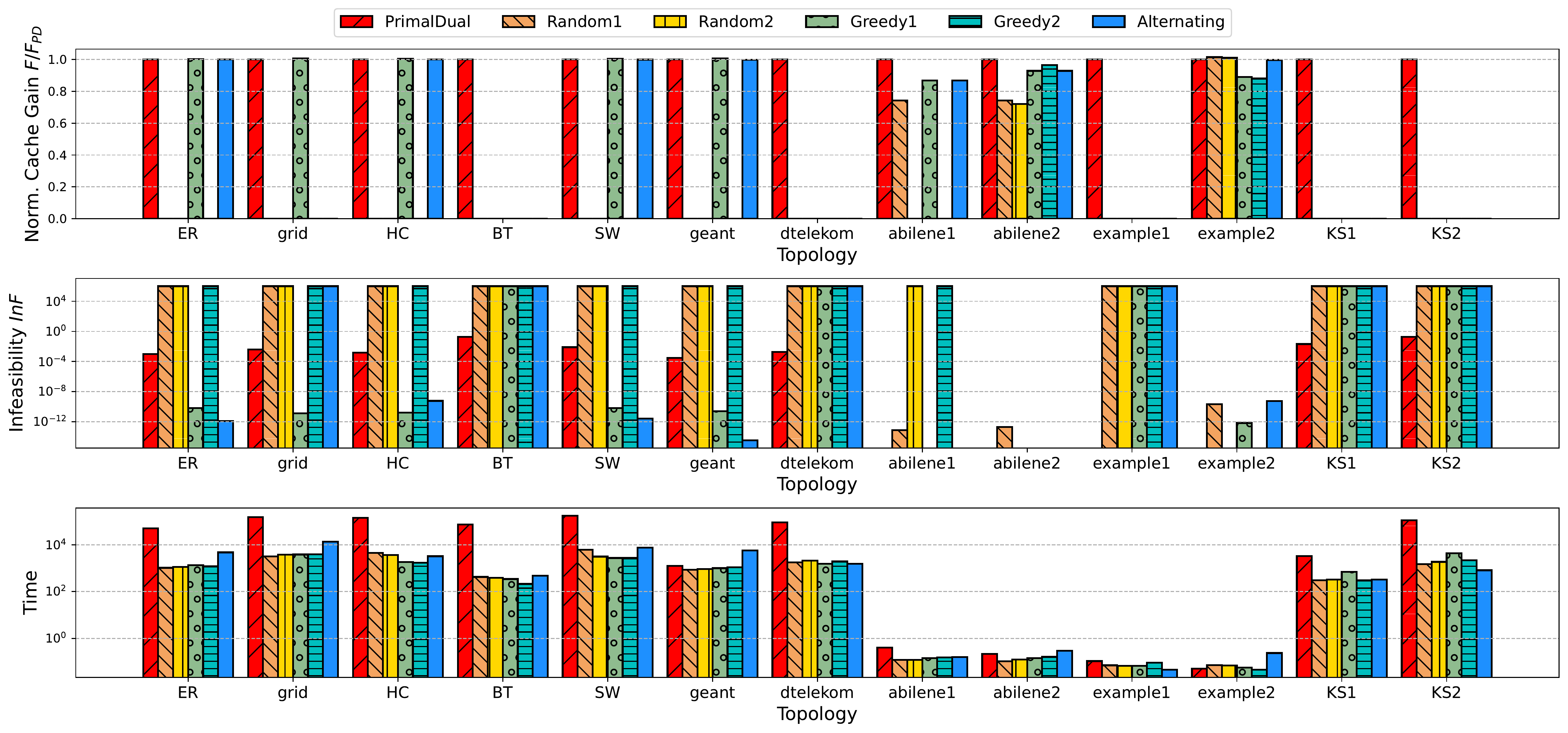}}
    \subcaptionbox{Looseness coefficient $\kappa=3$\label{fig:topologies3}}{\includegraphics[width = 0.9\linewidth]{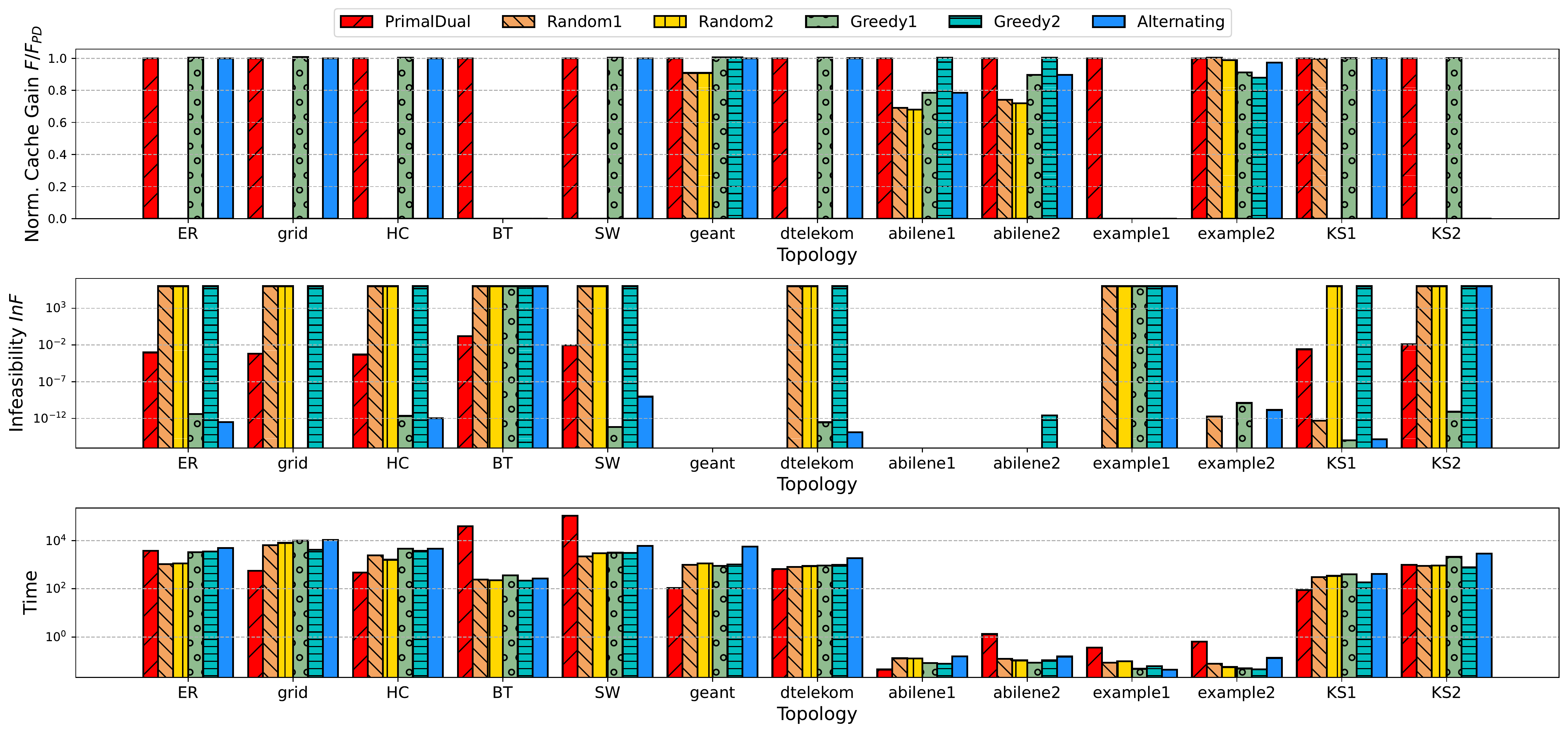}}
    \caption{Comparison w.r.t. gain, $\texttt{InF}$ \notes{and running time}. Missing bars in cache gain plots indicate infeasibility, while missing bars in $\texttt{InF}$ plots indicate 0 violation. \texttt{PrimalDual} and competitors perform very well when they are feasible. Nevertheless, \texttt{PrimalDual} is always feasible for all topologies, while competitors fail to get a feasible solution in some cases. \notes{This comes at the cost of the increased complexity of \texttt{PrimalDual}, reflected also on the running time. However, when $\kappa$ is large, i.e., $\kappa=3$, as \texttt{PrimalDual} converge faster, it takes even less execution time compared to competitors.}}
    \label{fig:topologies}
\end{figure*}

\begin{figure*}[t]
    \centering
    \subcaptionbox{with momentum and $\kappa=1$ \label{fig:converge1}}{\includegraphics[width = 0.24\linewidth]{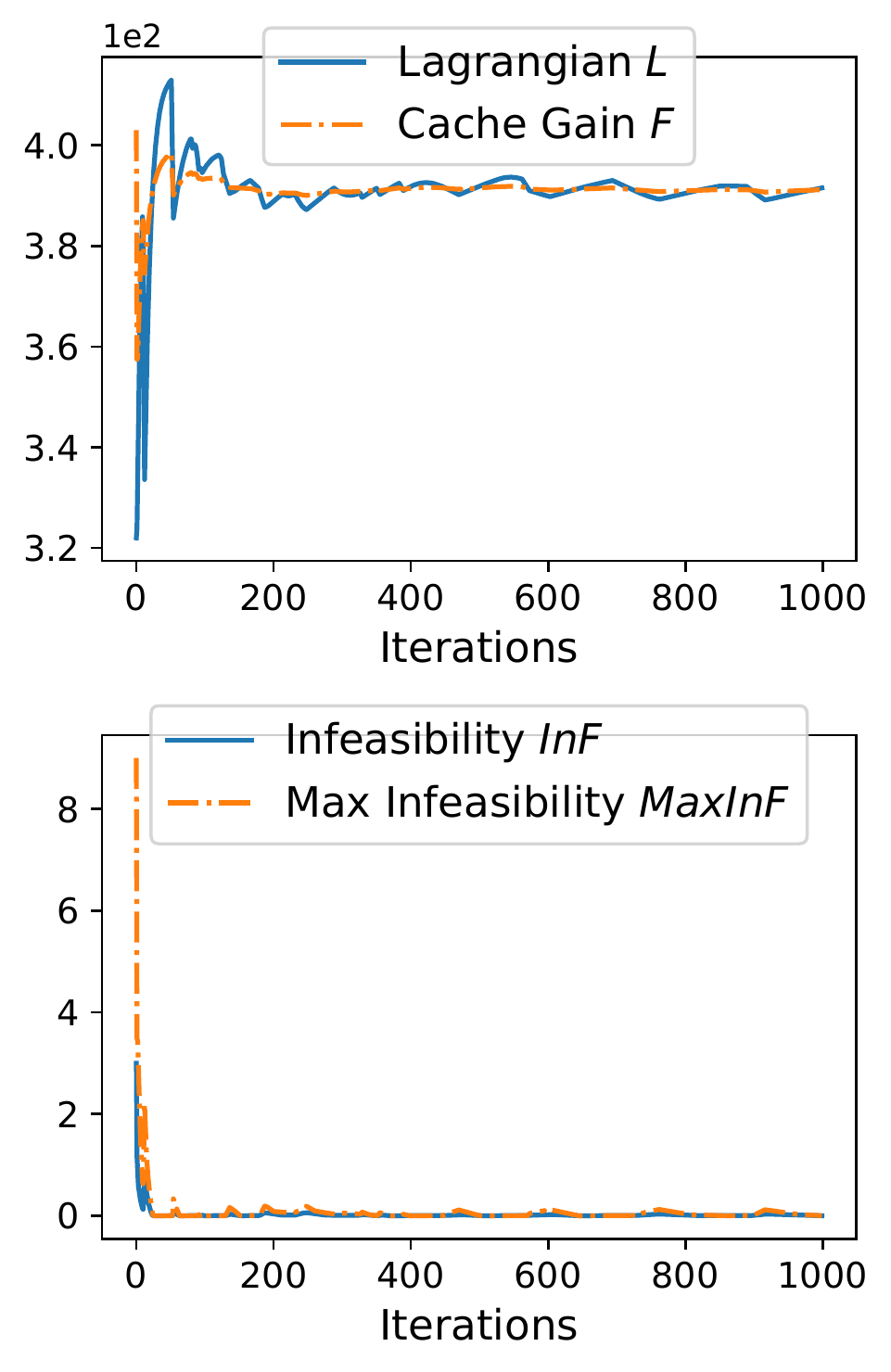}}
    \subcaptionbox{without momentum and $\kappa=1$\label{fig:converge2}}{\includegraphics[width = 0.24\linewidth]{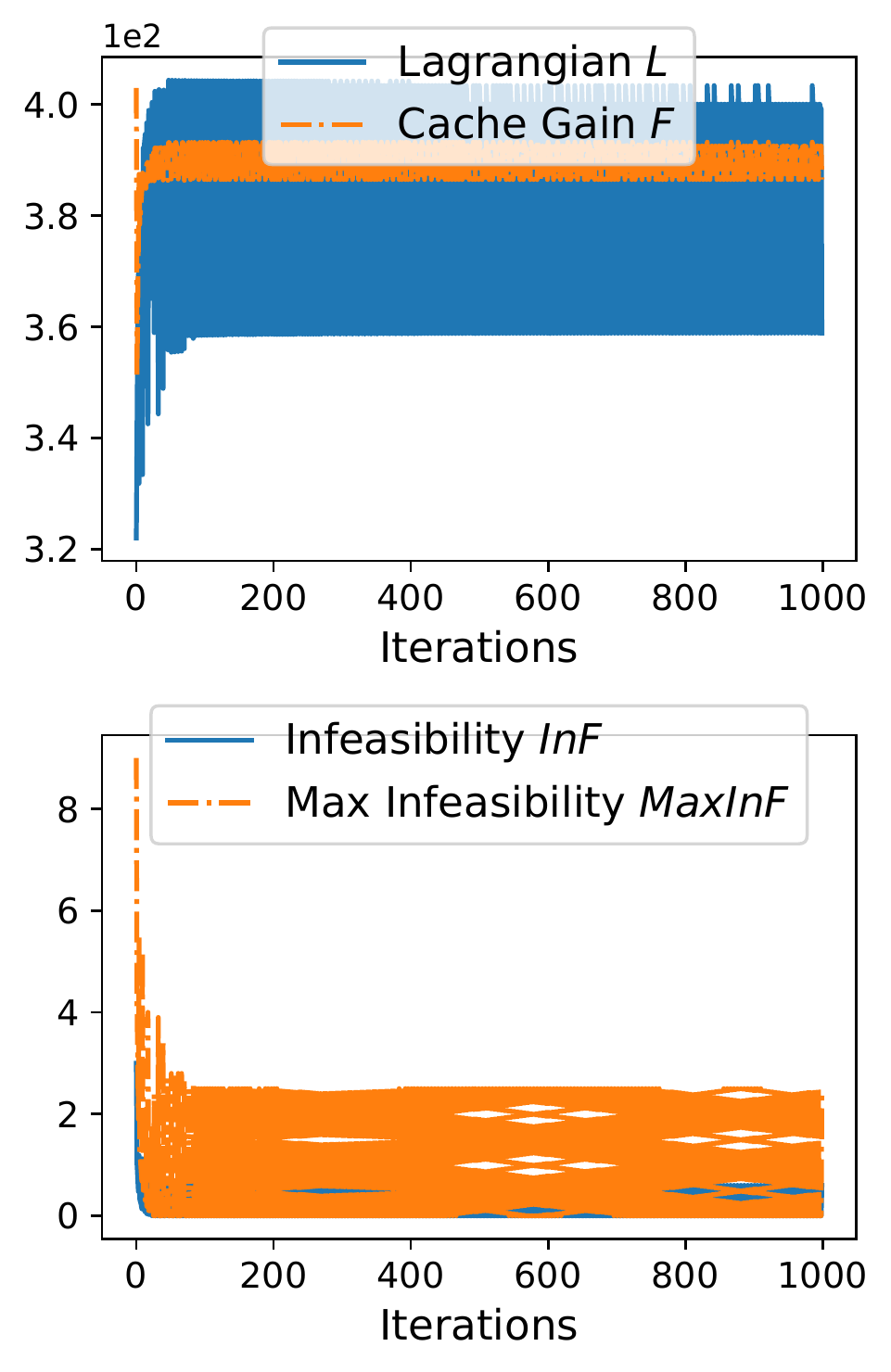}}
    \subcaptionbox{with momentum and $\kappa=3$\label{fig:converge3}}{\includegraphics[width = 0.24\linewidth]{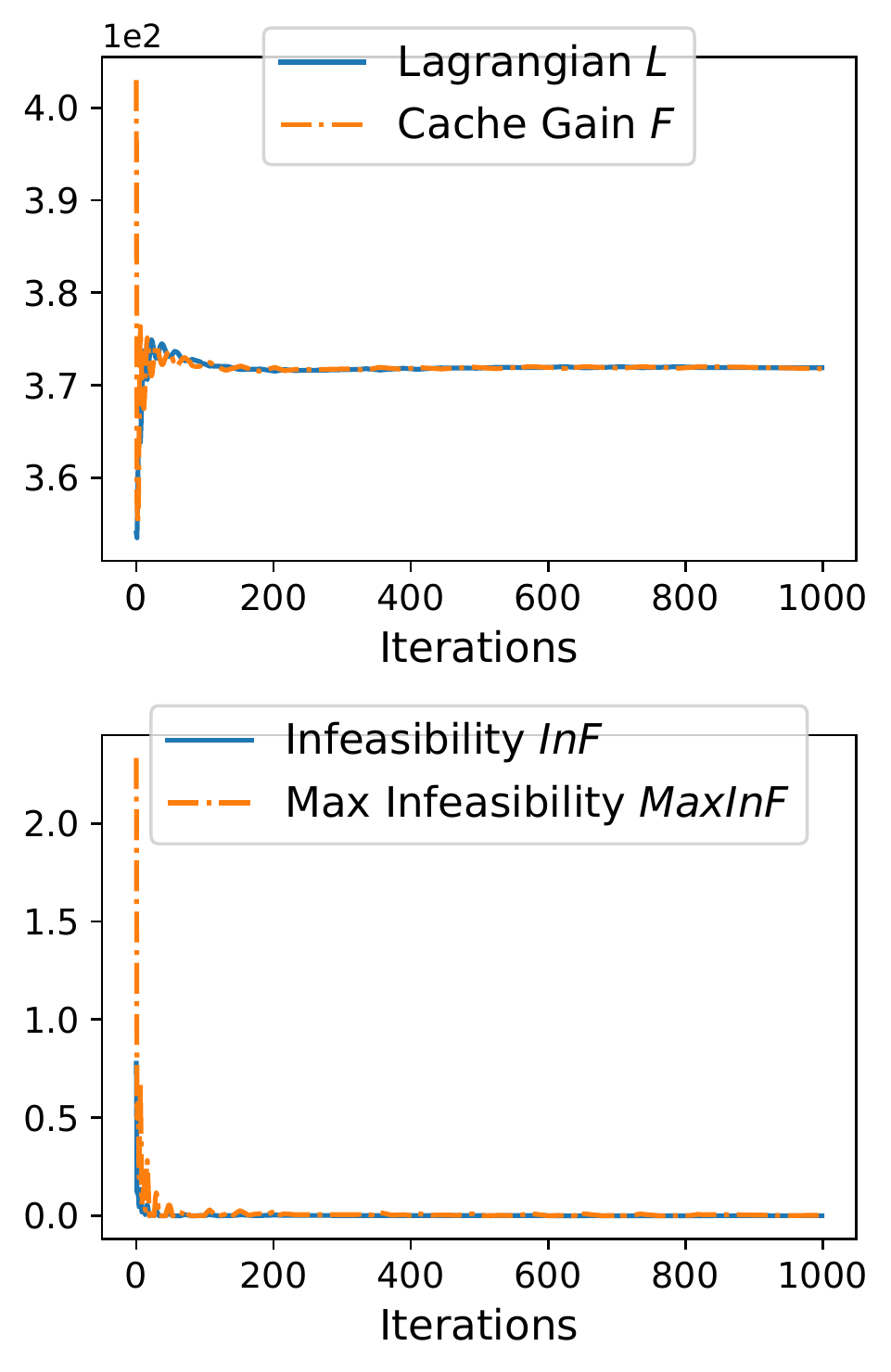}}
    \subcaptionbox{ without momentum and $\kappa=3$\label{fig:converge4}}{\includegraphics[width = 0.24\linewidth]{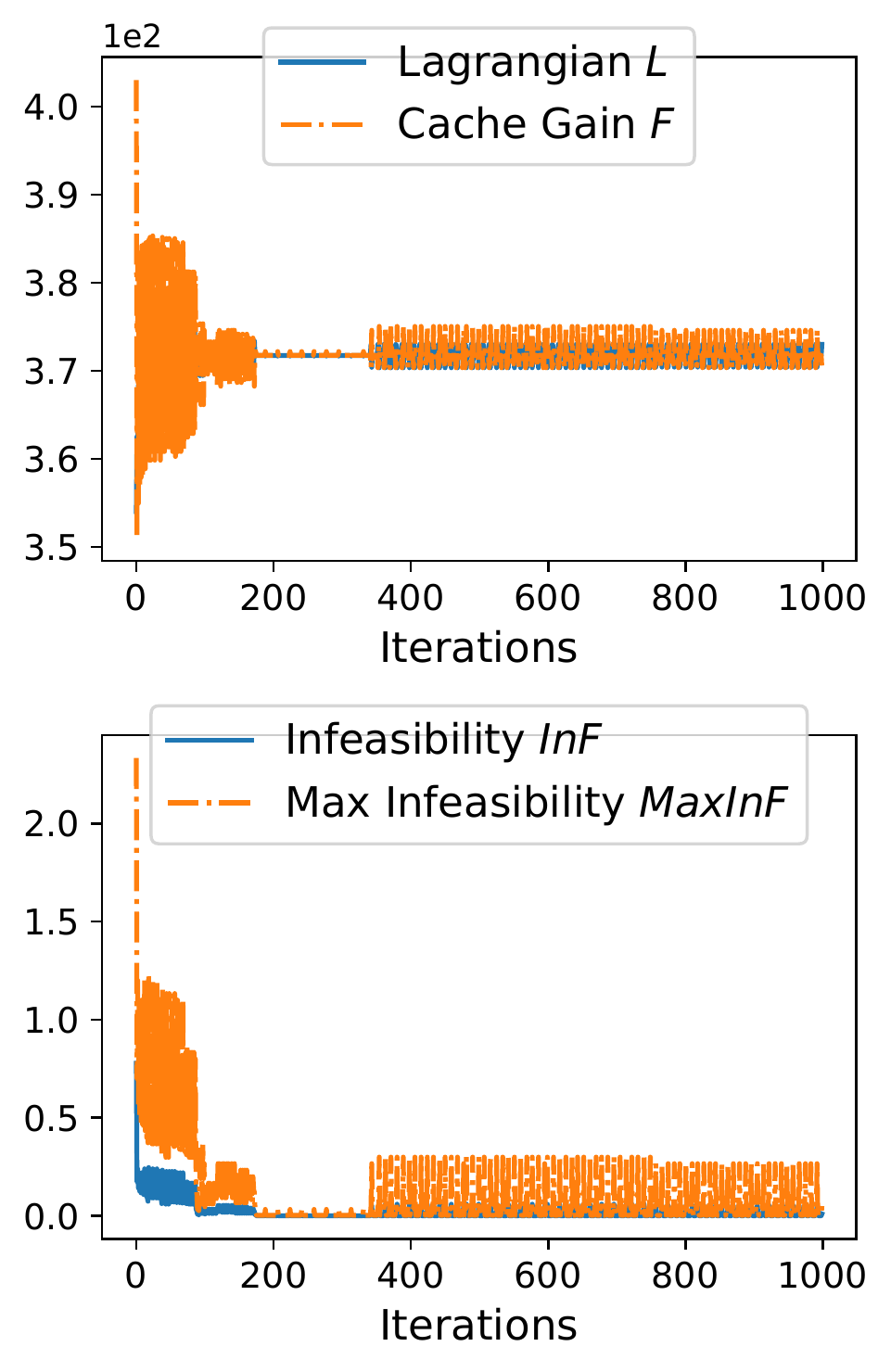}}
    \caption{Convergence over topology \texttt{Ex1}, w.r.t.~the cache gain, the Lagrangian $L$, and infeasibility metrics $\texttt{InF}$ and $MaxInf$. With momentum, algorithm converges faster and smoother.}
    \label{fig:convergence}
\end{figure*}
}{}

\subsection{Algorithms}\label{sec:Algorithms} We implement our algorithm and several competitors\footnote{Our implementation is publicly available at \href{https://github.com/neu-spiral/CacheRateNetwork}{https://github.com/neu-spiral/CacheRateNetwork}.} for comparison purposes. Our main building blocks when constructing competitors are combinations of algorithms that make caching and routing decisions separately. 

In particular, building blocks for caching are: (a) \emph{uniform caching}, whereby cache contents are selected uniformly among requests that traverse the cache, (b) \emph{greedy caching}, whereby the greedy algorithm \cite{krause2014submodular} is used to allocate items to caches, and (c) \emph{Frank-Wolfe variant caching}, where the Frank-Wolfe variant algorithm \cite{bian2017guaranteed} is used to determine cache contents; all three variants (a)--(c) are classic, but \emph{ignore edge capacity constraints}. The classic greedy algorithm \arxiv{starts from empty caches and makes  placements incrementally that maximizes objective \eqref{eq:cache_gain} subject only to cache capacity constraints. This caching decision }{}is a 1/2 approximation \cite{krause2014submodular, mahdian2020kelly}\arxiv{ \emph{if one ignores the edge capacity constraints}. The Frank-Wolfe}{, while the Frank-Wolfe} variant \cite{bian2017guaranteed} \arxiv{that maximizes objective \eqref{eq:cache_gain} subject to constraints \eqref{eq:down closed convex}, i.e., ignoring routing constraints, in a manner siminar to Alg.~\ref{alg:FWvariant}. This is a 1-1/e approximation algorithm if one ignores the edge capacity constraints, as the corresponding problem is DR-submodular maximization over down-closed convex set}{ achieves $1-1/e$}.
We combine these caching algorithms with \emph{optimal routing}, which amounts to fixing a caching strategy\arxiv{ (computed via uniform caching, greedy, etc.)}{}, and computing routing decisions by solving Prob.~\eqref{maxgain} w.r.t. routing decisions alone; this is a convex optimization problem with affine constraints, and can be solved in polynomial time. 

Overall, we implement the following combinations of these building blocks: 
\arxiv{
\begin{itemize}
    \item \texttt{Random1} consists of two steps. First, we assume all paths are active, and use uniform caching: we select caching decisions by placing items  in a cache selected u.a.r. from requests that traverse it. Having made caching decisions this way, we then set routing variables via optimal routing. Formally, in Step 1,  we first initialize $\tilde{\routeprobvec} = \boldsymbol{0}$, and then $\cacheprob_{v,i} = \min \{ \frac{\capacity_v}{\sum_{\overset{i \in \catalog: v \in p,}{p \in \pathset_{(i,s)}}} 1}, 1 \}$, for all $v\in V, i \in \catalog$. In Step 2, keeping $\cacheprobvec$ fixed, we optimize the Prob.~\eqref{maxgain} w.r.t.  routing variables~$\tilde{\routeprobvec}$.
    \item \texttt{Random2} also consists of two steps. In Step 1, we use optimal routing assuming empty caches: that is, we solve Prob.~\eqref{maxgain} w.r.t. the routing variables $\tilde{\routeprobvec}$ assuming $\cacheprobvec = \boldsymbol{0}$. In Step 2, we again fix $\tilde{\routeprobvec}$ as computed from the previous step, and determine $\cacheprobvec$ via uniform caching, as in the first step of \texttt{Random1}. 
    \item \texttt{Greedy1} consists of the following two steps. In Step~1, we initialize $\tilde{\routeprobvec} = \boldsymbol{0}$, and then use greedy caching, i.e., make caching decisions using the classic greedy algorithm \cite{krause2014submodular}. 
    In Step 2, having $\cacheprobvec$ from greedy caching, we determine routing variables $\tilde{\routeprobvec}$ via optimal routing.
    \item \texttt{Greedy2} also consists of two steps; in Step 1, we initialize $\cacheprobvec = \boldsymbol{0}$ (i.e., empty caches), and determine routing variables $\tilde{\routeprobvec}$ via optimal routing. In Step~2, fixing $\tilde{\routeprobvec}$ from Step~1, we determine $\cacheprobvec$ via greedy caching.
    \item \texttt{Alternating} solves Prob.~\eqref{maxgain} via alternating maximization between caching and routing variables,  until convergence. It first initializes $\tilde{\routeprobvec} = \boldsymbol{0}$, and then updates caching decisions $\cacheprobvec$ and routing decisions $\tilde{\routeprobvec}$ alternately. When updating caching decisions $\cacheprobvec$, we fix $\tilde{\routeprobvec}$ and determine $\cacheprobvec$ through the Frank-Wolfe variant \cite{bian2017guaranteed}; that is, maximize objective \eqref{eq:cache_gain} subject to constraints \eqref{eq:down closed convex}, i.e., ignoring routing constraints, through the Frank-Wolfe variant \cite{bian2017guaranteed}. When updating routing decisions $\tilde{\routeprobvec}$, we fix $\cacheprobvec$ and determine the new $\tilde{\routeprobvec}$ via optimal routing. We repeat this process for at most 25 iterations (we observe experimentally that \texttt{Alternating} converges within 10 iterations).
    \item \texttt{PrimalDual} is our algorithm (Algorithm \ref{alg:PrimalDual}).We set the number of iterations to 1000 steps.
\end{itemize} 
}{\texttt{Random1}, that uses uniform caching first  and then optimal routing;  \texttt{Random2}, that performs  optimal routing under empty caches first, and then performs uniform caching; \texttt{Greedy1}, that uses greedy caching first and then optimal routing;  \texttt{Greedy2}, that uses optimal routing  under empty caches first, and then greedy caching; \texttt{Alternating}, that alternates (until convergence) between obtaining a caching state via the Frank-Wolfe variant algorithm and optimal routing; and \texttt{PrimalDual} is demonstrated in \eqref{eq:pdalgo}. }
\arxiv{We discuss convergence criteria in Section~\ref{sec: metrics}. We implement all algorithms in Python, and use the \texttt{CVXPY} toolbox to solve constituent convex optimization problems (e.g., during optimal routing). Overall, the above competitors  decompose the problem into two subproblems: determining caching decisions $\cacheprobvec$ and routing decisions $\tilde{\routeprobvec}$, and edge capacities are taken into account in the latter optimization. This decoupling may lead to infeasibility; we prove this formally in Appendix \ref{sec:proof suboptimal}, where we construct several counterexamples under which the above algorithms lead to solutions violating edge capacity constraints and experimentally in Sec.~\ref{sec:experiment results}.} {Additional implementation details are described in \cite{li2022extended}. In \cite{li2022extended}, we prove that \emph{all of the above competitors} (\texttt{Random1}--\texttt{Alternating}) can lead to arbitrarily suboptimal solutions.  \addtolength{\topmargin}{0.04in}
} 

\arxiv{}{
\begin{figure}[t]
    \centering
    \subcaptionbox{Looseness coefficient $\kappa=1$\label{fig:topologies1}}{\includegraphics[width = 1.\linewidth]{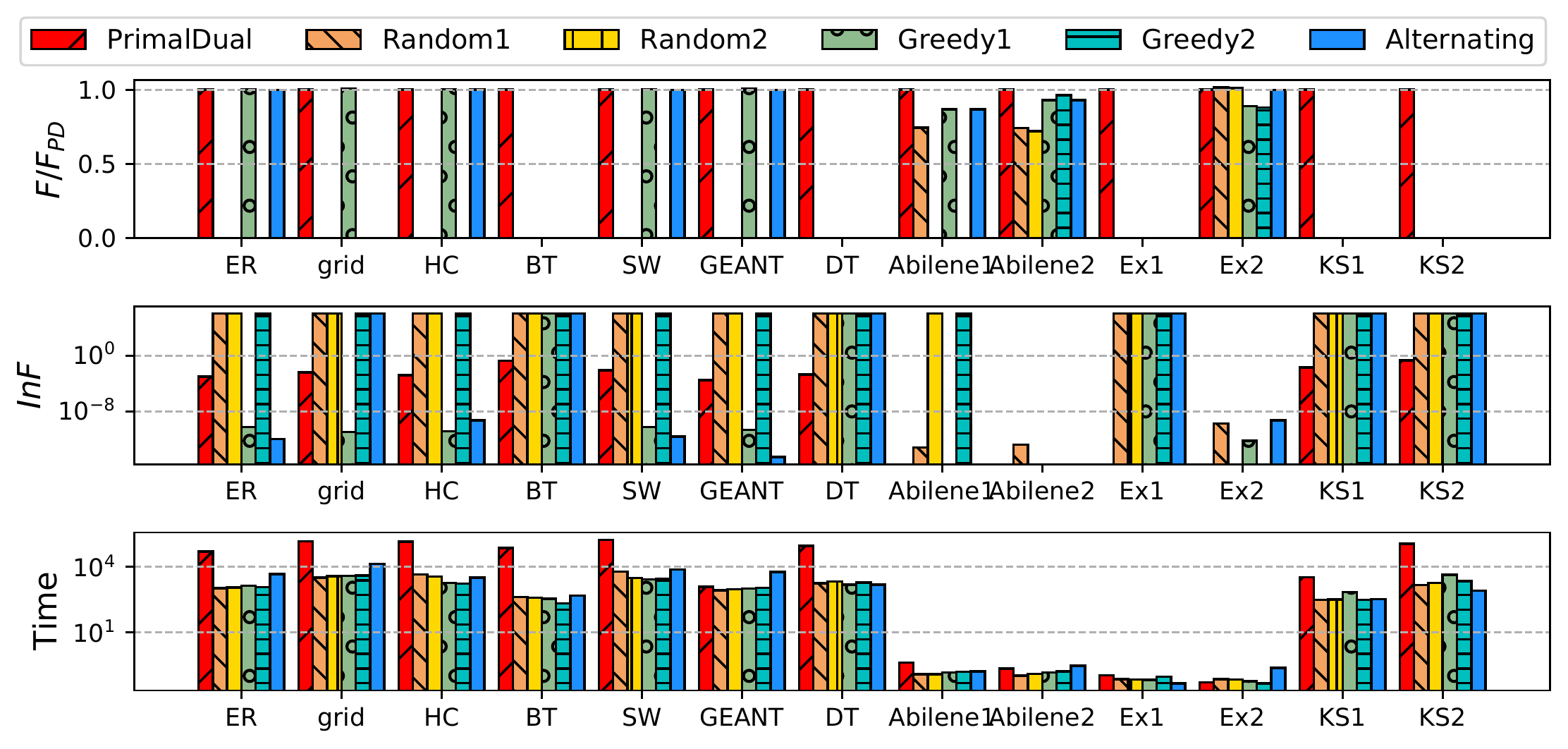}}
    \subcaptionbox{Looseness coefficient $\kappa=3$\label{fig:topologies3}}{\includegraphics[width = 1.\linewidth]{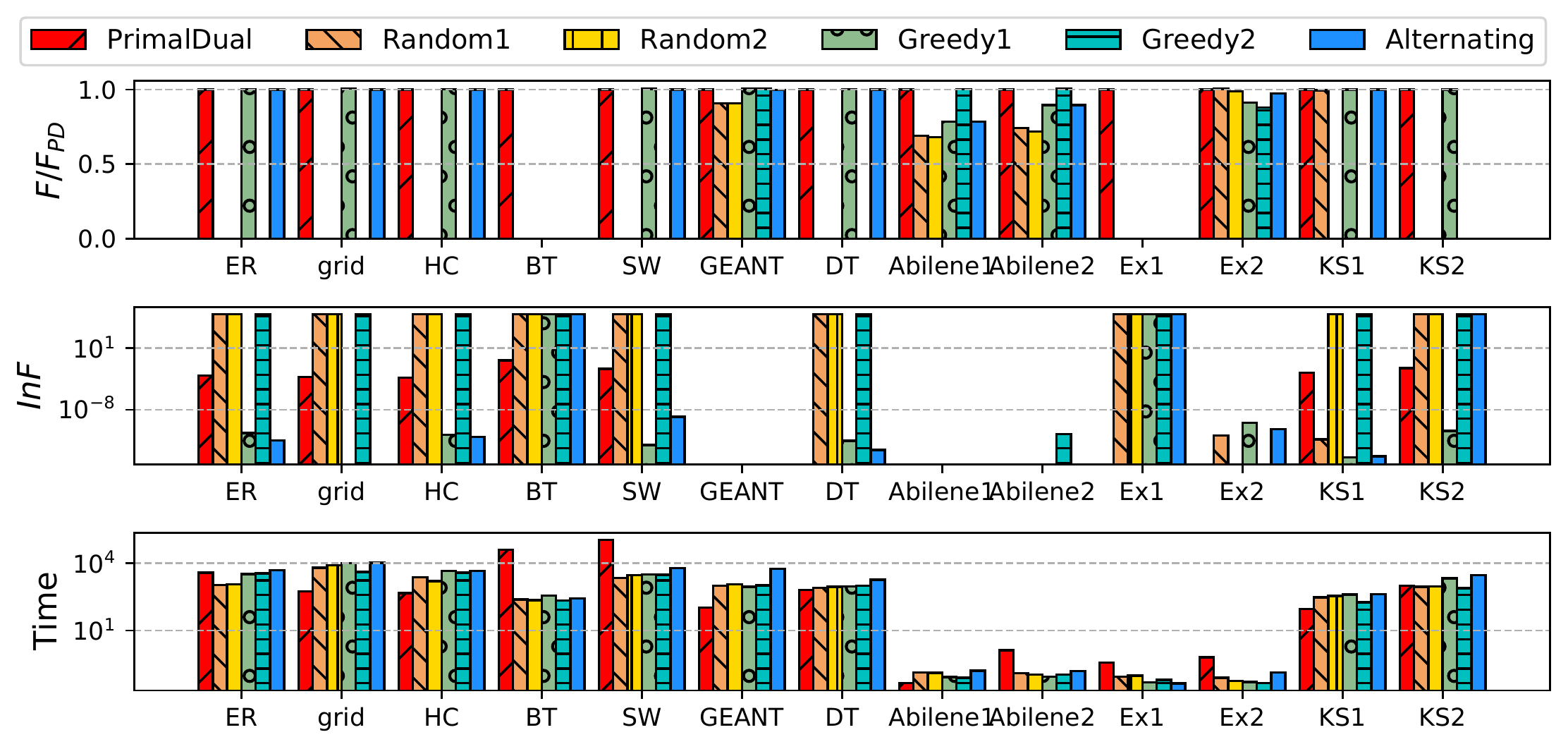}}
    \caption{Comparison w.r.t. gain, $\texttt{InF}$ \notes{and running time}. Missing bars in cache gain plots indicate infeasibility, while missing bars in $\texttt{InF}$ plots indicate 0 violation. \texttt{PrimalDual} and competitors perform very well when they are feasible. Nevertheless, \texttt{PrimalDual} is always feasible for all topologies, while competitors fail to get a feasible solution in some cases. \notes{This comes at the cost of the increased complexity of \texttt{PrimalDual}, reflected also on the running time. However, when $\kappa$ is large, i.e., $\kappa=3$, as \texttt{PrimalDual} converge faster, it takes even less execution time compared to competitors.}}
    \label{fig:topologies}
\end{figure}

\begin{figure}[t]
    \centering
    \subcaptionbox{with momentum and $\kappa=1$ \label{fig:converge1}}{\includegraphics[width = 0.49\linewidth]{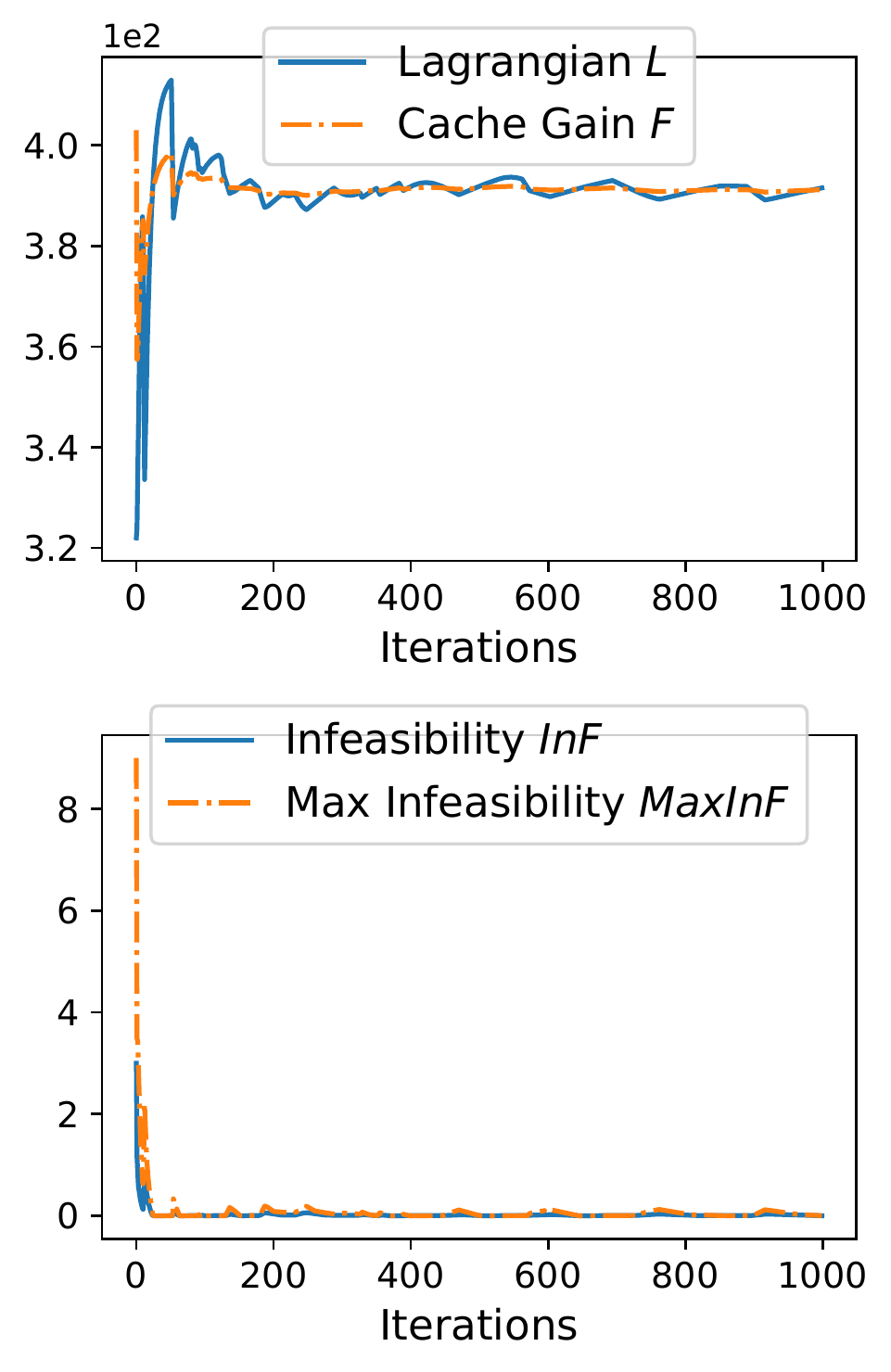}}
    \subcaptionbox{without momentum and $\kappa=1$\label{fig:converge2}}{\includegraphics[width = 0.49\linewidth]{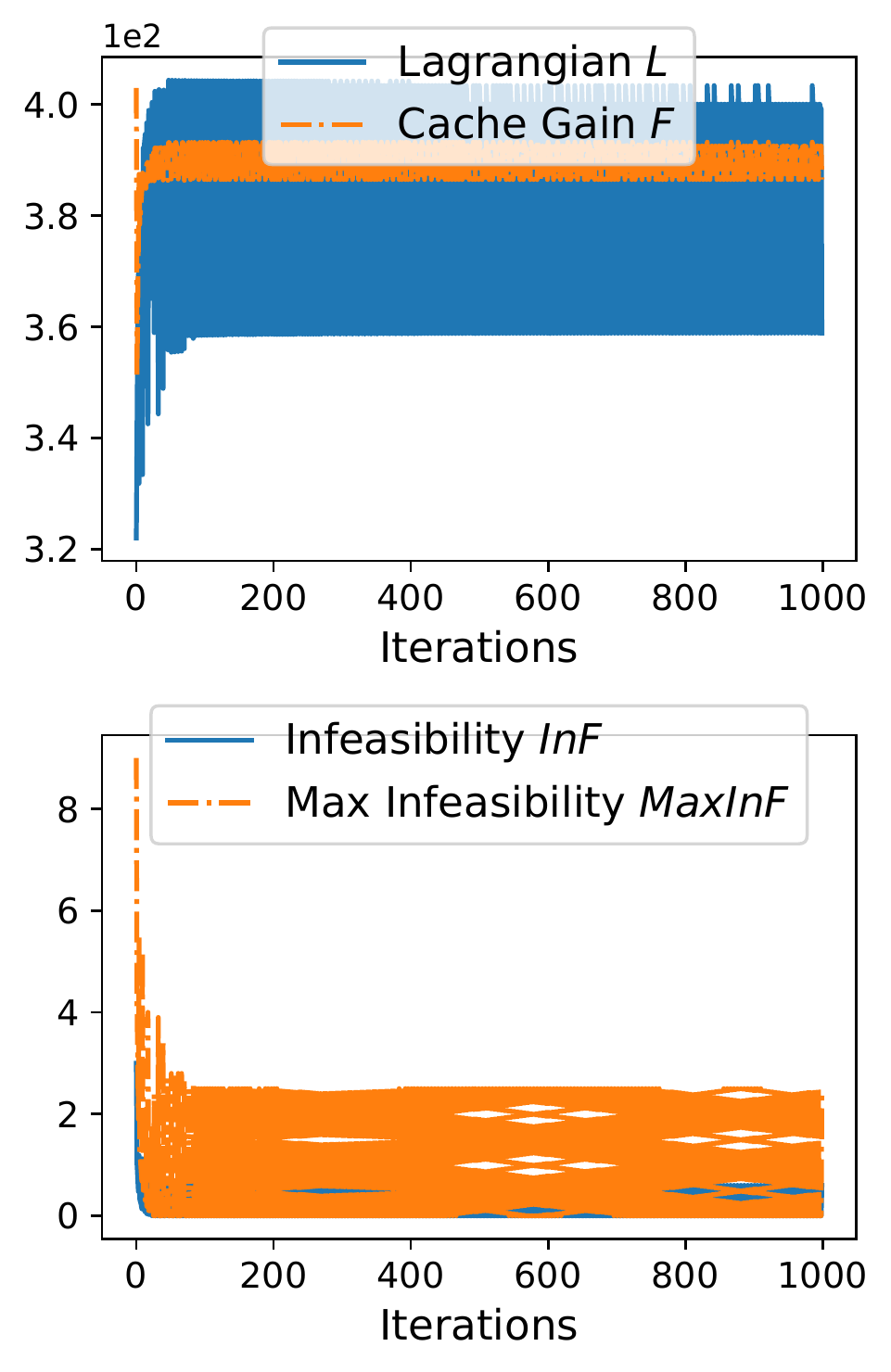}}
    \caption{Convergence over topology \texttt{Ex1}, w.r.t.~the cache gain, the Lagrangian $L$, and infeasibility metrics \texttt{InF} and \texttt{MaxInf}. With momentum, algorithm converges faster and smoother.}
    \label{fig:convergence}
\end{figure}
}

\subsection{Performance Metrics}
\label{sec: metrics}
We use cache gain, defined in Eq.~\eqref{eq:cache_gain}, as one metric to measure the performance of different algorithms. Also, we define an \emph{Infeasibility} metric to measure how much solutions violate link capacity constraints. Intuitively, we measure infeasibility as the average overflow, normalized by edge capacities, across all active edges in the network. Formally:
\begin{equation}
    \texttt{InF} = \frac{1}{|E'|} \sum_{(u,v)\in E'} \frac{G_{u,v}(\cacheprobvec,\tilde{\routeprobvec}) \id_{G_{u,v}(\cacheprobvec,\tilde{\routeprobvec})>0}}{\mu_{u,v}},
\end{equation}
where overflow $G_{u,v}(\cacheprobvec,\tilde{\routeprobvec})$ is defined in Eq. \eqref{eq:overflow}, and $E'=\{e\in E: \lambda_{e}((\cacheprobvec,\tilde{\routeprobvec})>0\}$ is the set of edges with non-zero flow, and flow $\lambda_{e}$ is defined in Eq.~\eqref{eq:flow}. \final{We say algorithms \texttt{Alternating} and \texttt{PD} algorithm converge, when $\texttt{InF}\le 0.001$ and cache gain changes less than 0.001 compared to the last iteration. }We also report \texttt{MaxInF}, which is the maximum rather than average over $E'$\arxiv{, i.e., 
\begin{equation}
    \texttt{MaxInF} = \max_{(u,v)\in E'} \{ \frac{G_{u,v}(\cacheprobvec,\tilde{\routeprobvec}) \id_{G_{u,v}(\cacheprobvec,\tilde{\routeprobvec})>0}}{\mu_{u,v}}\},
\end{equation}}{.}
Clearly, larger $\texttt{InF}/\texttt{MaxInf}$ indicates more violations and worse performance. For our algorithm \texttt{PrimalDual}, we expect some negligible edge capacity constraint violation, of the order of $\texttt{InF}\sim 10^{-2}$. Whenever \texttt{CVXOPT} fails to find a feasible solution, we are unable to compute this score\arxiv{ (as no $\tilde{\routeprobvec}$ is provided to evaluate this)}{}, so  we set $\texttt{InF}=10^6$, to indicate a severe feasibility failure. 

\arxiv{
\begin{figure*}[t]
    \centering
\subcaptionbox{\texttt{Abilene2}\label{fig:looseness1}}{\includegraphics[width = 0.3\linewidth]{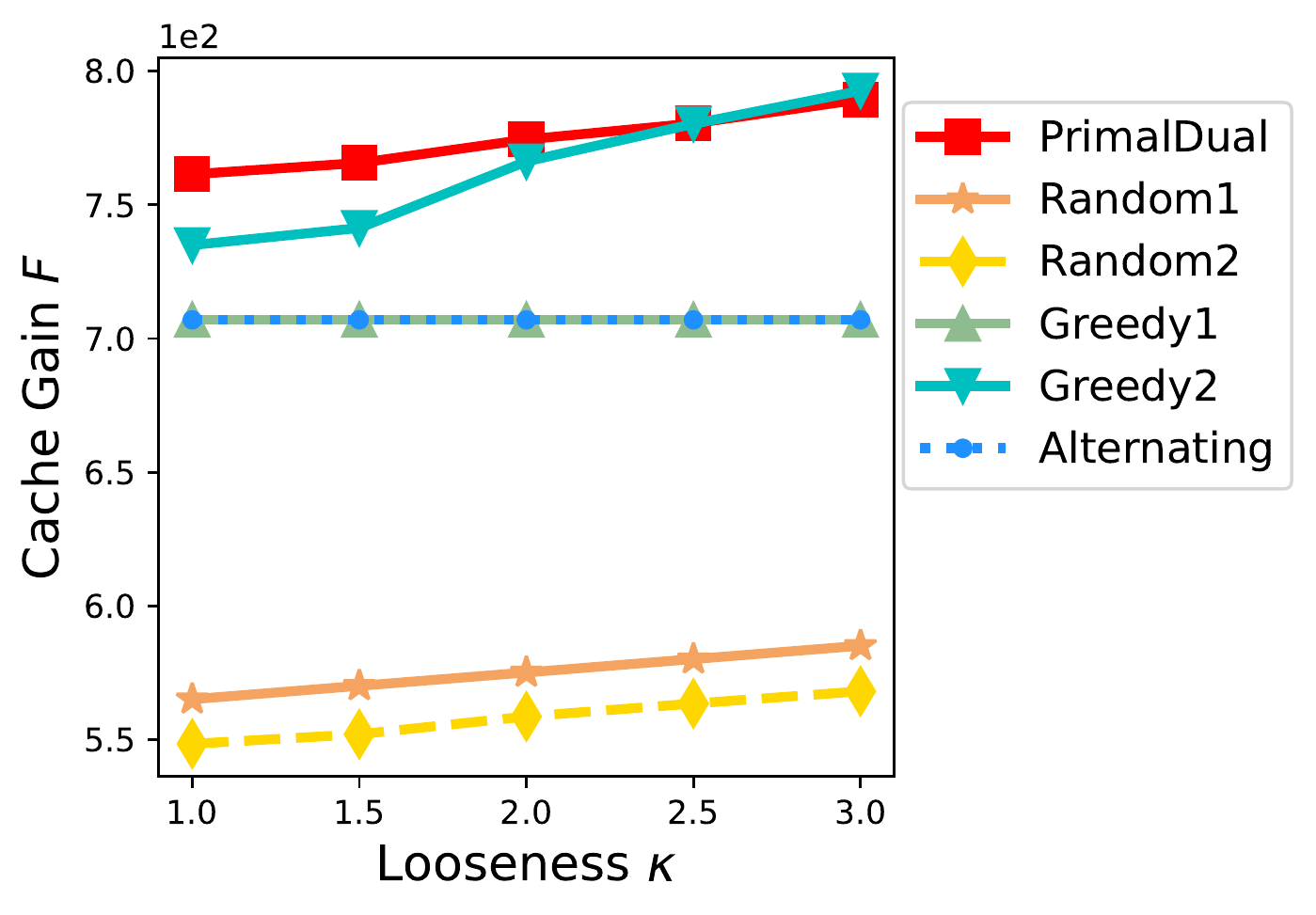}}
    \subcaptionbox{\texttt{GEANT}}{\includegraphics[width = 0.3\linewidth]{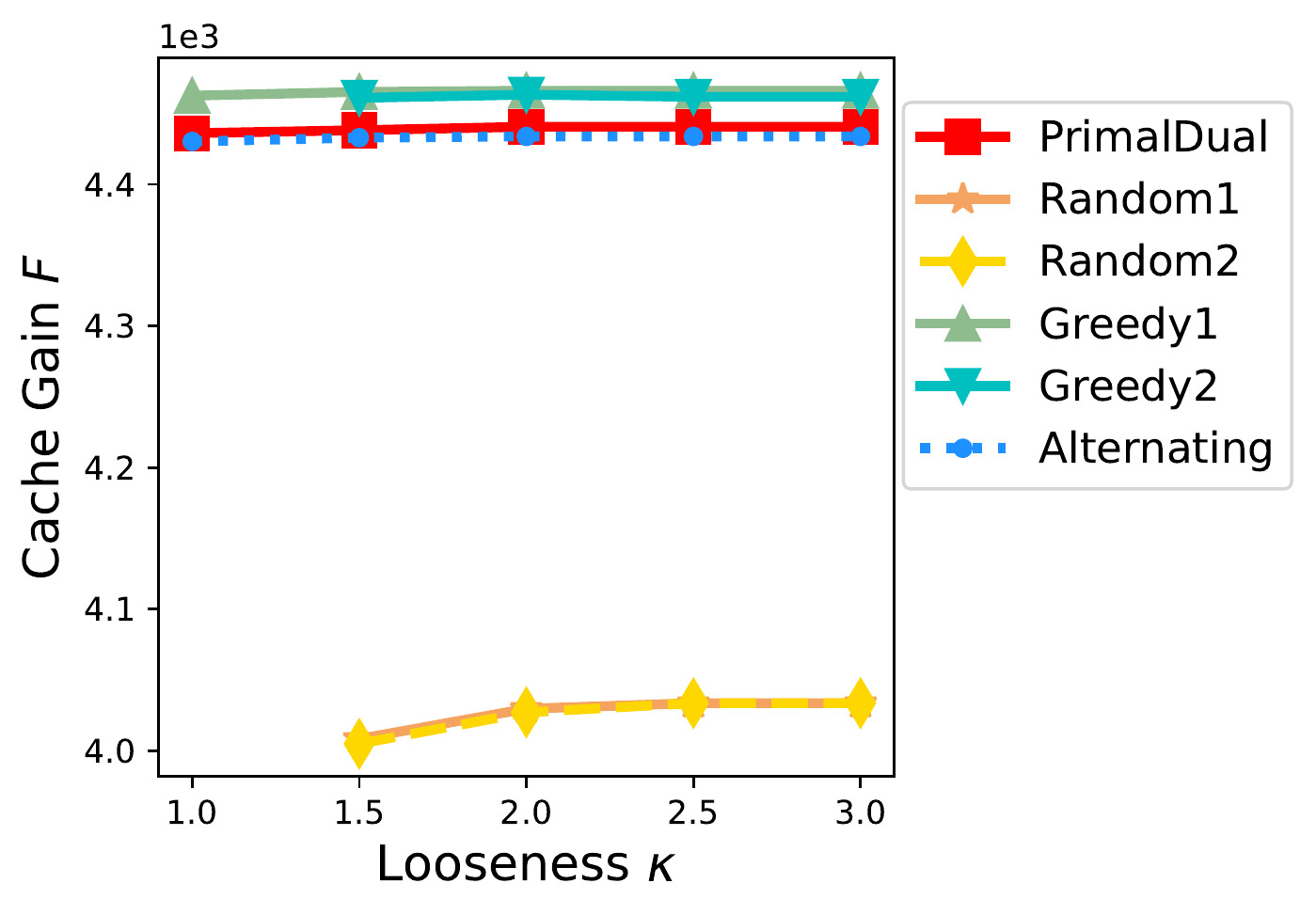}}
    \subcaptionbox{\texttt{KS1}}{\includegraphics[width = 0.3\linewidth]{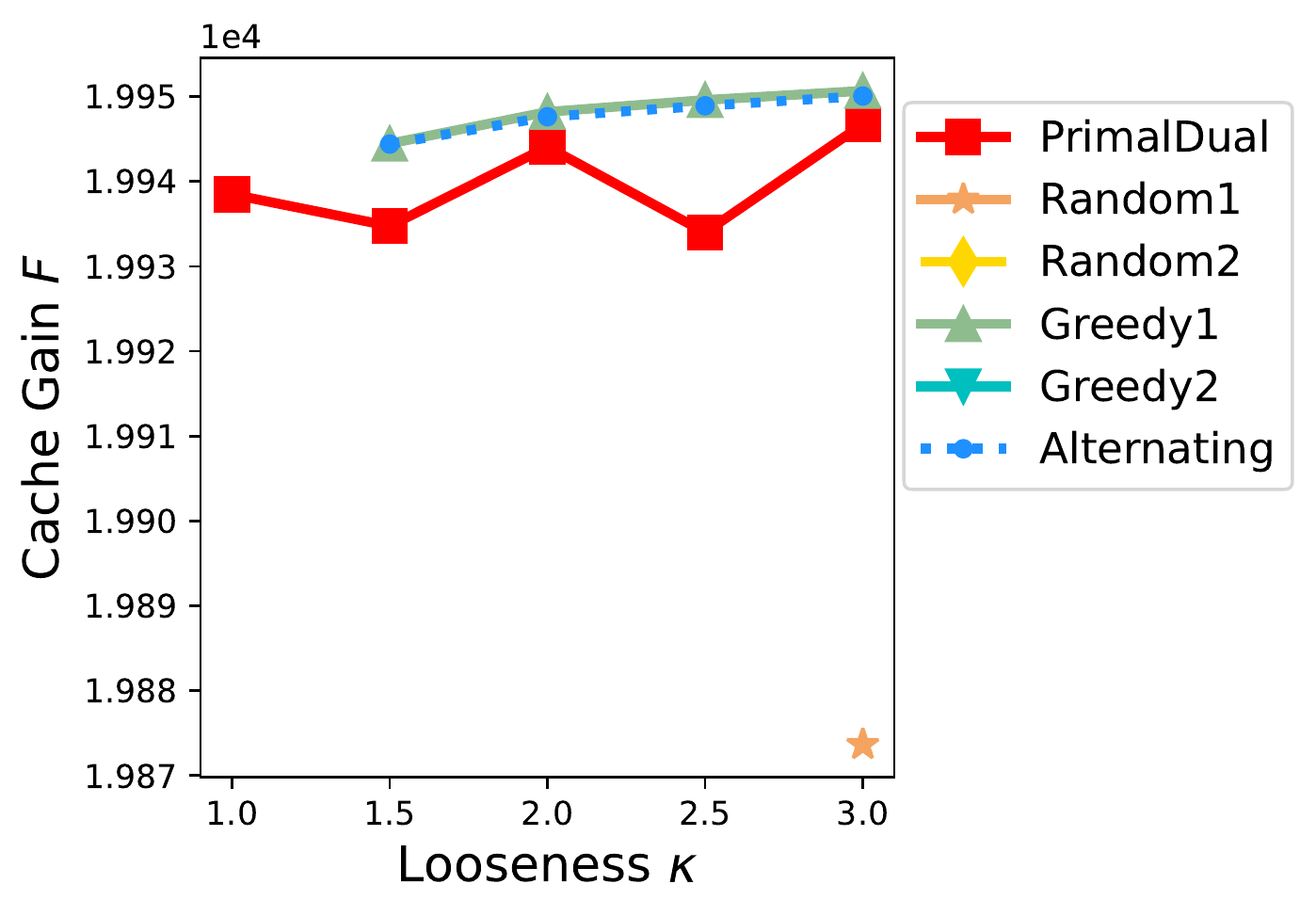}}
    \caption{Effect of looseness. \texttt{PrimalDual} is best or competitive with other algorithms, but also finds a feasible solution for a wider range of $\kappa$ values. }
    \label{fig:looseness}
\end{figure*}
}{}

\subsection{Experiment Results}
\label{sec:experiment results}

\noindent\textbf{Different Topologies. }
We first compare the proposed algorithm (\texttt{PrimalDual}) with baselines in terms of the normalized cache gain $\frac{F}{F_{\texttt{PD}}}$, Infeasibility $\texttt{InF}$, \notes{and running time of algorithms, }shown in Fig.~\ref{fig:topologies}. The cache gain $F_{\texttt{PD}}$ is obtained by \texttt{PrimalDual} algorithm, and its value is reported in Tab.~\ref{tab:setting}: $F_{\texttt{PD}}^1$ is the cache gain when $\kappa=1$, while $F_{\texttt{PD}}^3$ when $\kappa=3$. When algorithms obtain no feasible solutions, normalized cache gain and infeasibility are set 0 and $10^6$, correspondingly. Observe that \texttt{PrimalDual}, \texttt{Greedy1}, \texttt{Greedy2} and \texttt{Alternating} behave great w.r.t. cache gain when they are feasible. However, even though \texttt{PrimalDual} \emph{always produces a feasible solution}  (with $\texttt{InF}\sim 10^{-2}$ consistently), solutions of other algorithms are infeasible in some topologies. This is because \texttt{PrimalDual} jointly optimizes both caching and routing decisions. In other words, in every intermediate step, it takes link capacity constraints into consideration. In contrast, competitors decouple routing and caching optimization,  ignoring link capacity constraints in the latter. This verifies the suboptimality of competitors. \notes{These advantages of \texttt{PrimalDual} come at the cost of increased running time; nevertheless, with larger looseness $\kappa$, \texttt{PrimalDual} converges faster, sometimes even outperforming simpler methods.}

\noindent\textbf{Convergence. }
\label{sec:convergence}
We focus on \texttt{Ex1} to understand the convergence of proposed \texttt{PrimalDual}. \arxiv{\notes{Instead of terminating the algorithm based on convergence, we execute the algorithm for $1000$ iterations. }}{} \arxiv{Figs.~\ref{fig:converge1} and~\ref{fig:converge3} demonstrate}{Fig.~\ref{fig:converge1} demonstrates} the convergence with momentum (defined in Eq. \eqref{eq: primal}). Both cache gain and infeasibility converge smoothly and quickly. On the other hand, without momentum, i.e., for $\alpha_t = 1$, both cache gain and infeasibility exhibit jitter, as shown in \arxiv{Figs.}{Fig.}~\ref{fig:converge2}\arxiv{ and \ref{fig:converge4}}. \arxiv{Compared to momentum, algorithms}{Algorithms} without momentum tend to converge to a more infeasible solution. Overall, incorporating momentum in primal steps avoids oscillations in primal variables and promotes faster and smoother convergence.
\arxiv{
}{
\begin{figure}[t]
    \centering
    \includegraphics[width = 1\linewidth]{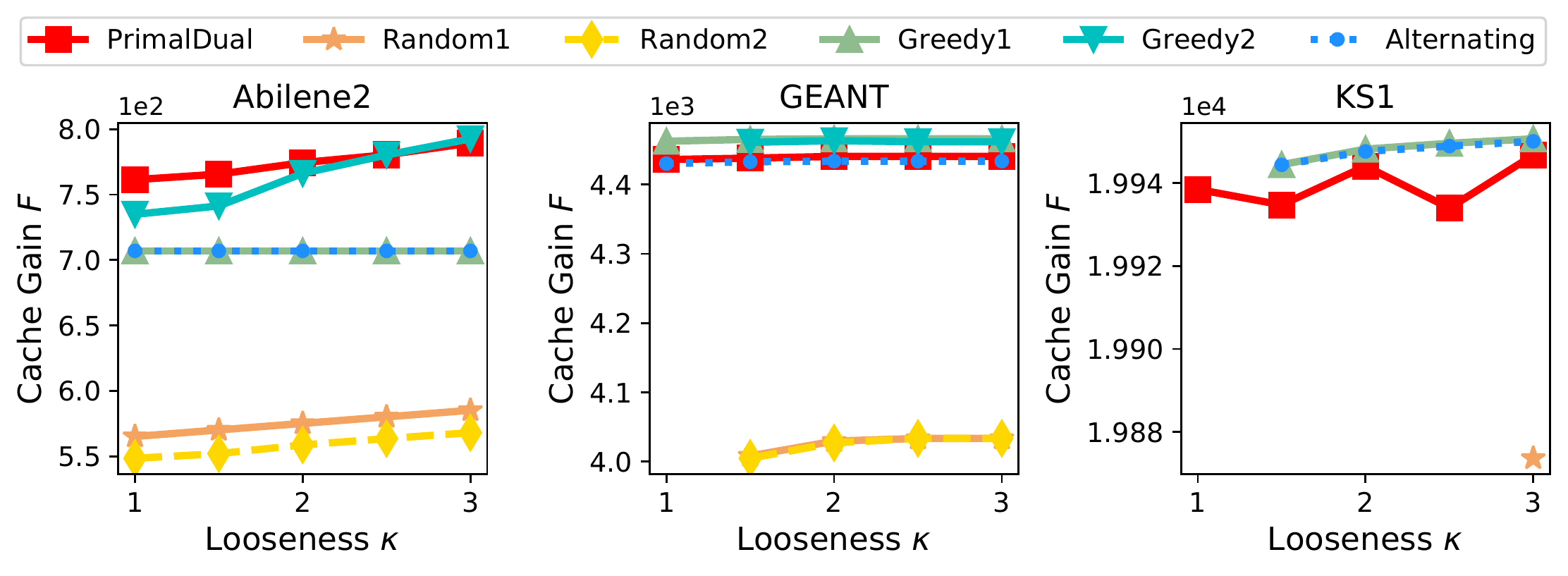}
    \caption{Effect of looseness over \texttt{Abilene2}, \texttt{GEANT} and \texttt{KS1}, respectively. \texttt{PrimalDual} is best or competitive with other algorithms, but also finds a feasible solution for a wider range of $\kappa$ values.}
    \label{fig:looseness}
\end{figure}
}

\noindent\textbf{Effect of Looseness. }
Figs. \ref{fig:topologies}, \arxiv{\ref{fig:convergence}, }{}\ref{fig:looseness} and Tab. \ref{tab:setting} all present results with different looseness coefficient $\kappa$.  When looseness $\kappa$ is small, i.e., link capacity constraints are strict and hard to satisfy, competitors are more likely to lead to infeasibility. It is clear from Fig. \ref{fig:topologies} and Tab. \ref{tab:setting} that, in general, higher $\kappa$ leads to \arxiv{higher cache gain and less infeasibility}{higher cache gain, less infeasibility \notes{and faster convergence/less execution time}}. This is also indicated directly in Fig.~\ref{fig:looseness}: if algorithms have no results at some $\kappa$, this indicates infeasibility. In contrast, although not always obtaining the highest cache gain, our proposed \texttt{PrimalDual} \emph{always yields a solution}, and is near-optimal. \arxiv{In Fig.~\ref{fig:convergence} \notes{and running time in Fig. \ref{fig:topologies}}, we observe that higher $\kappa$ results in lower infeasibility (see y-axis), and faster convergence \notes{ (less execution time)}.}{}

\section{Conclusion}
\label{sec:conclusion}
We jointly optimize both caching and routing decisions under bounded link capacity constraints over an arbitrary network. We propose a poly-time primal-dual algorithm, where only primal steps have an approximation guarantee. We use a momentum method to alleviate sharp changes in primal variables. Instead, we could explore a proximal method \cite{bolte2014proximal,bitterlich2019proximal} to realize it. As we only provide approximation guarantees for primal steps, another direct and crucial future direction is to propose an algorithm with end-to-end optimality guarantees. 

\bibliographystyle{IEEEtran}  
\bibliography{references}

\arxiv{
\appendices
\section{Proof of Theorem \ref{thm:FW variant}}
\label{sec:proof FW variant}
\begin{proof}
Frank-Wolfe variant algorithm shown in Alg.~\ref{alg:FWvariant} is a classic method \cite{bian2017guaranteed} for:
\begin{equation}
    \max_{\jvar\in \feasibledomain' } L(\jvar,\dualvec),
\end{equation}
which is a continuous DR-submodular maximization problem under down-closed convex constraint.
We first prove that constraints $\feasibledomain'$ are binding, i.e., there exists an optimal point $y^{**} = \argmax_{\jvar\in \feasibledomain' } L(\jvar,\dualvec)$, such that the inequality \eqref{pselect_2} in $\feasibledomain'$:
\begin{equation}
\label{pselect_4}
    \sum_{p\in \pathset_{(i,s)}} \tilde{\routeprob}_{(i,s),p} \le |\pathset_{(i,s)}|-1, \text{ for all }(i,s)\in\requests,
\end{equation}
holds with equality \eqref{pselect_1} in $\feasibledomain$, i.e.:
\begin{equation}
\label{pselect_3}
    \sum_{p\in \pathset_{(i,s)}} \tilde{\routeprob}_{(i,s),p} =|\pathset_{(i,s)}|-1, \text{ for all }(i,s)\in\requests,
\end{equation}
hence, $\jvar^{**}\in\feasibledomain$.
Suppose that equality \eqref{pselect_3} does not hold for any optima $\jvar^{**}$, i.e., $\sum_{p\in \pathset_{(i,s)}} \tilde{\routeprob}_{(i,s),p} < |\pathset_{(i,s)}|-1$, for some $(i,s)\in\requests$. Hence, there must exist a $\jvar' > \jvar^{**}$ (at least for one coordinate $\jvar'_i > \jvar^{**}_i$), s.t. $\jvar' \in \feasibledomain'$, while also $\jvar' \in \feasibledomain$ (i.e., constraints bind). By the monotonicity of $L$,  we would then have $L(\jvar',\dualvec) \geq L(\jvar^{**},\dualvec)$. Hence, $\jvar'\in \feasibledomain$ is also an optimum (in $\feasibledomain'$).
This binding indicates that there exists a $\jvar^{**} = \argmax_{\jvar\in \feasibledomain' } L(\jvar,\dualvec)$ also being a solution to \eqref{MaxSubmodular}. As an optimal solution to problem \eqref{MaxSubmodular}, $\jvar^*$ implies $L(\jvar^{*},\dualvec) \ge L(\jvar^{**},\dualvec)$. Furthermore, feasible set $\feasibledomain'$ is larger than $\feasibledomain$ because of \eqref{pselect_2}, thus $L(\jvar^{**},\dualvec) \ge L(\jvar^{*},\dualvec)$. To sum it up, $L(\jvar^{**},\dualvec) = L(\jvar^{*},\dualvec)$. 

Similarly, because of monotonicity of $\langle \boldsymbol{v}, \nabla L(\jvar,\dualvec(t)) \rangle$, there exists an optima $\boldsymbol{v}_k$, such that \eqref{pselect_2} in $\feasibledomain'$ holds with equality \eqref{pselect_1} in $\feasibledomain$. Thus, $\jvar_{\texttt{FW}} = \sum_k \gamma_k\boldsymbol{v}_k$, where $\sum_k \gamma_k = 1$, as a convex combination of points in $\feasibledomain$, also in $\feasibledomain$.

\notes{
According to Bian et al. \cite{bian2017guaranteed}, the Frank-Wolfe variant algorithm has the following performance guarantee:
\begin{lemma}
\label{lem:bian}
For a non-negative DR-submodular continuous function $f$ maximization problem under down-closed convex constraint, a fixed number of iterations K, and constant stepsize $\gamma_k = \gamma = K^{-1}$, Alg. \ref{alg:FWvariant} provides the following approximation guarantee:
\begin{equation}
 f(\jvar_K) -  f(\boldsymbol{0}) \ge (1-\frac{1}{e})\left(f(\jvar^{**}) - f(\boldsymbol{0})\right) - \frac{M}{2K} ,\label{eq:final}
\end{equation}
where $\jvar_K$ is the output of Alg. \ref{alg:FWvariant}.
\end{lemma}
}
Lagrangian $L$, defined by Eq. \eqref{eq:lagrangian}, could attain negative values. To provide an optimality factor, we offset $L$ by a constant; i.e., let $f(\jvar)\equiv L(\jvar,\dualvec)+C$, where \notes{$C = \sum_{e\in E} \dual_e (\sum_{(i,s)\in \requests} \allowbreak \lambda_{(i,s)}-\mu_e)$ is an upper bound on $\sum_{e\in E} \dual_{u,v} \cdot G_{u,v}(\cacheprobvec,\tilde{\routeprobvec})$}. Then,  $f$ is DR-submodular, non-negative, and    $f(\boldsymbol{0}) = 0$. We thus have:
\begin{equation}
\begin{split}
    L(\jvar_{\text{FW}},\dualvec)+C \overset{\text{Lemma}~\ref{lem:bian}}{\ge} (1-\frac{1}{e}) (L(\jvar^{**},\dualvec)+C) \notes{-\frac{M}{2K}} \\ =
    (1-\frac{1}{e}) (L(\jvar^{*},\dualvec)+C) \notes{-\frac{M}{2K}},
\end{split}
\end{equation}
\notes{where $M = 2 L(\boldsymbol{1},\dualvec)(|V||\catalog|+P_\SR)^2$ is the Lipschitz continuous
constant, and the theorem follows.  
}
\end{proof}

\section{Proof of suboptimality of competitors}
\label{sec:proof suboptimal}
In \texttt{example1} and $\kappa = 1$, the optimal caching decision is $\cacheprob_{4,\text{blue}} = 1$, $\cacheprob_{5,\text{blue}} = 1$, $\cacheprob_{3,\text{orange}} = 1$ and else equal to 0.
\begin{itemize}
    \item \texttt{Random1} generates $\cacheprob_{4,\text{blue}} = \cacheprob_{4,\text{orange}} = 0.5$, $\cacheprob_{5,\text{blue}} = 1$, $\cacheprob_{3,\text{orange}} = 1$ and else equal to 0 in Step 1. Thus, no $\tilde{\routeprobvec}$ can satisfy link capacities of edges $(2,4)$ and $(5,7)$.
    \item \texttt{Random2} has no feasible solution of determining routing variables $\tilde{\routeprobvec}$ in Step 1.
    \item \texttt{Greedy1} generates $ \cacheprob_{4,\text{orange}} = 1$, $\cacheprob_{5,\text{blue}} = 1$, and else equal to 0 in Step 1. Similar to \texttt{Random1}, no $\tilde{\routeprobvec}$ can satisfy link capacities of edges $(2,4)$ and $(5,7)$.
    \item \texttt{Greedy2} encounters the same infeasibility as \texttt{Random2}.
    \item \texttt{Alternating} generates $\cacheprob_{4,\text{blue}} = \cacheprob_{4,\text{orange}} = 0.5$, $\cacheprob_{5,\text{blue}} = 1$, $\cacheprob_{3,\text{orange}} = 1$ and else equal to 0, when updating $\cacheprobvec$. Similar to \texttt{Random1}, no $\tilde{\routeprobvec}$ can satisfy link capacities of edges $(2,4)$ and $(5,7)$.
    \item \texttt{PrimalDual} generates a feasible solution with infeasibility $InF=0$ as shown in Figs. \ref{fig:converge1} and \ref{fig:topologies1}.
\end{itemize}
This example verifies the suboptimality of our competitors, although they perform pretty well when feasible.

From Fig. \ref{fig:topologies}, we see that \texttt{Greedy} has poor performance, compared with \texttt{Random}, while in all other cases, \texttt{Greedy} perform better.
From Figs. \ref{fig:topologies} and \ref{fig:looseness}, we see that in \texttt{Abilene}, our proposed \texttt{PrimalDual} outperforms competitors for any $\kappa$.

\begin{figure}[t]
    \centering
    \subcaptionbox{\texttt{Ex2} with $\kappa = 1$\label{fig:topology_example2}}{\includegraphics[width = 0.33\linewidth]{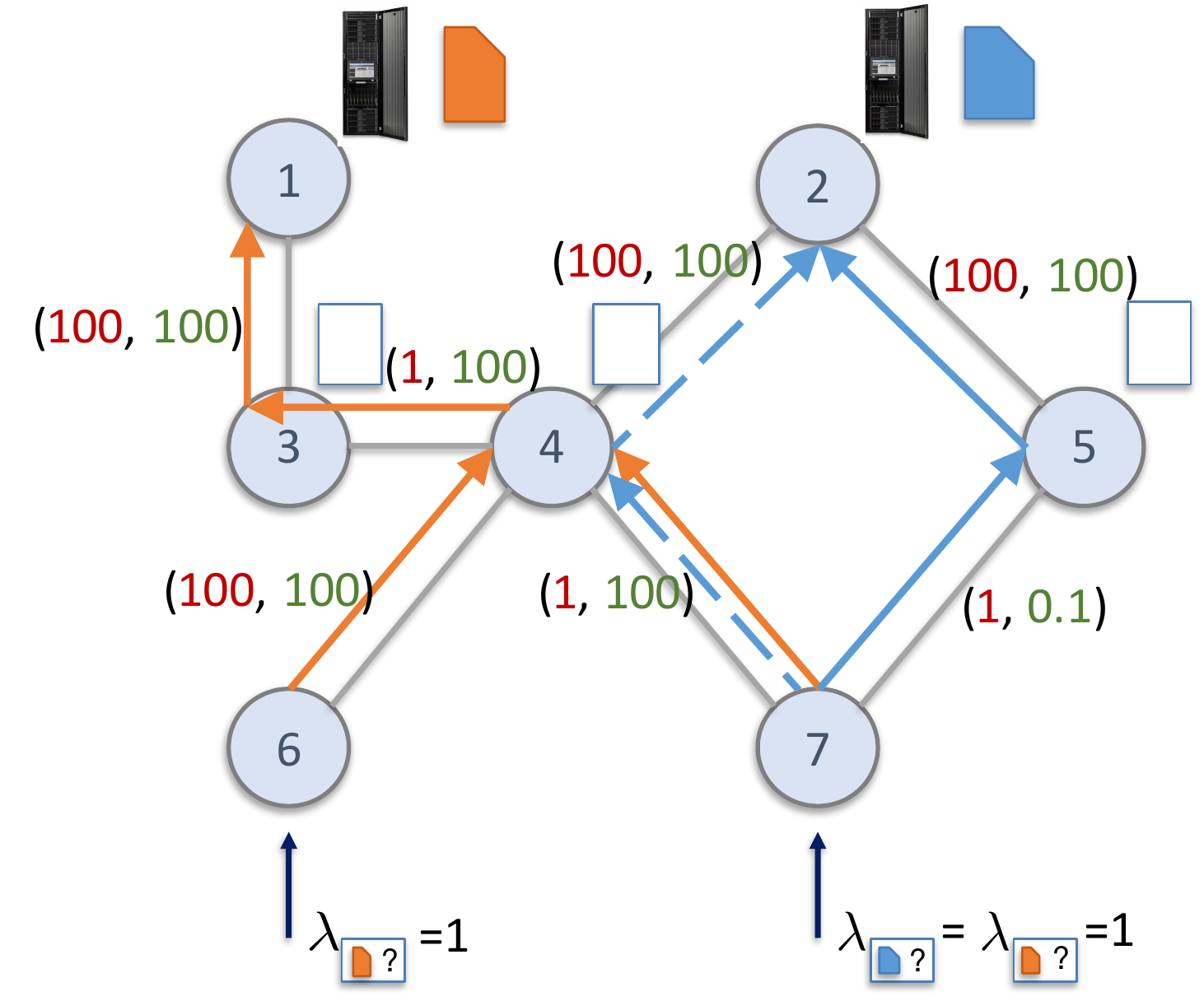}}
    \subcaptionbox{\texttt{Abilene2} with $\kappa = 1$\label{fig:topology_abilene2}}{\includegraphics[width = 0.63\linewidth]{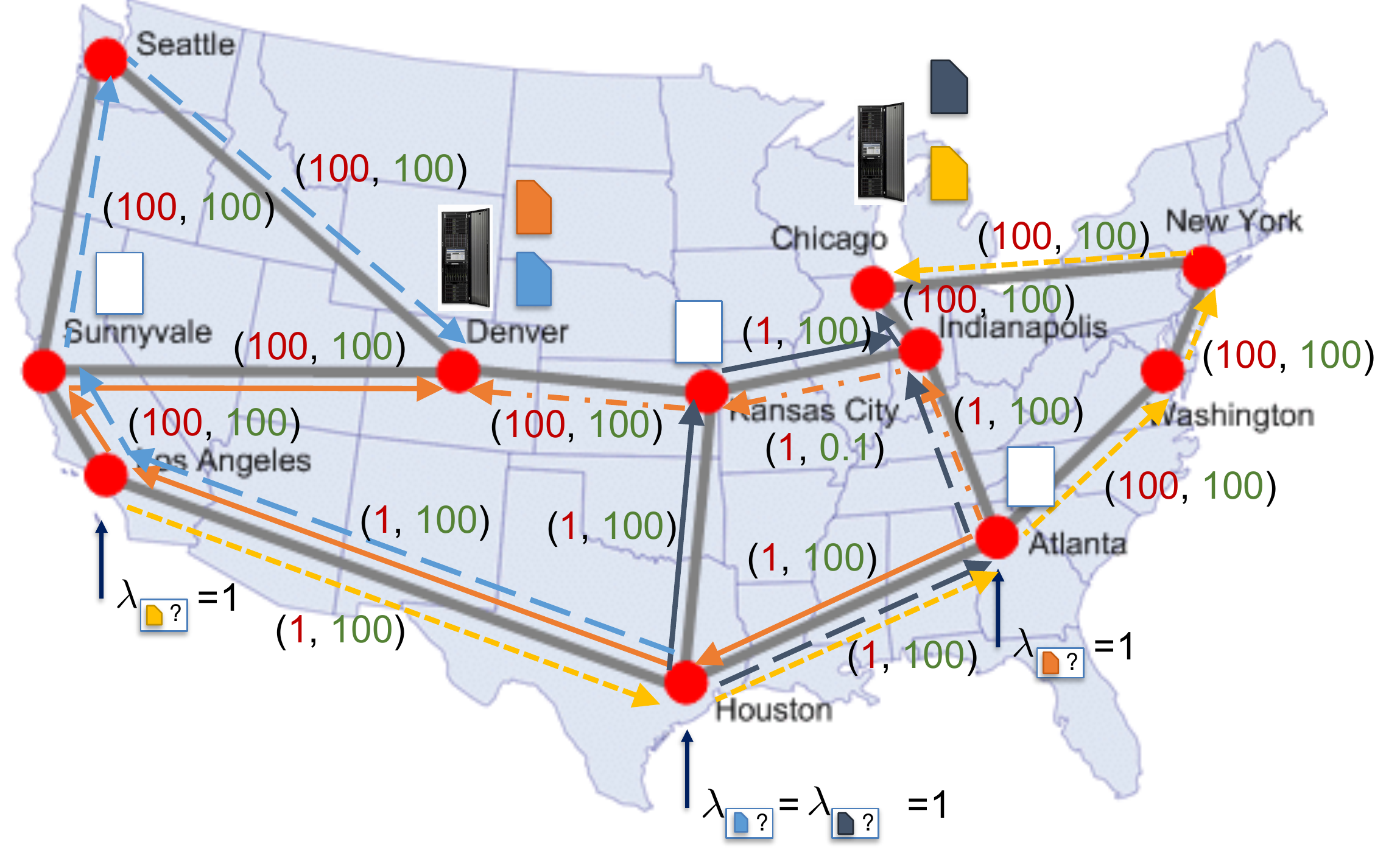}}
    \caption{Topologies and parameters of \texttt{Ex2} and \texttt{Abilene2} with designed requests and bandwidths. There is a pair (\textcolor{red}{\texttt{red}}, \textcolor[rgb]{0.1,0.6,0.3}{\texttt{green}}) for edge (u,v), where the first \textcolor{red}{\texttt{red}} number is the weight $w_{(v,u)}$ and the second \textcolor[rgb]{0.1,0.6,0.3}{\texttt{green}} number is the link capacity $\mu_{(v,u)}$.}
\end{figure}

\section{Parameters for Example and Abilene}
\label{sec:requests}
Parameter details are specified Fig.~\ref{fig:topology_example2} and Fig.~\ref{fig:topology_abilene2}, for \texttt{Ex2} and \texttt{Abilene2}, respectively. The differences between \texttt{Ex1} and \texttt{Ex2}, \texttt{Abilene1} and \texttt{Abilene2} are different link capacities for some of edges. In particular, \texttt{Ex1} and \texttt{Ex2} differ on edge $(2,4)$. This change is designed to make suboptimal algorithms infeasible in \texttt{Ex1} and have high cost in \texttt{Ex2}. \texttt{Abilene1} and \texttt{Abilene2} differ on edge (Indianapolis,Chicago), (Indianapolis,Kansas City), and (Sunnyvale to Denver). The differences are designed again so that the first topology is infeasible and the second leads to high cost under suboptimal algorithms.

}{}

\end{document}